\documentclass[a4paper,fleqn,usenatbib]{mnras}

\usepackage{newtxtext,newtxmath}
\usepackage[T1]{fontenc}
\usepackage{ae,aecompl}

\usepackage{graphicx}	% Including figure files
\usepackage{amsmath}	% Advanced maths commands
\usepackage{amssymb}	% Extra maths symbols

\title[CR-supported outflows]{Cooler and smoother -- the impact of cosmic rays on the phase structure of galactic outflows}

\author[Philipp Girichidis et al.]{Philipp~Girichidis$^{1,2,3}$\thanks{E-mail: philipp@girichidis.com}, Thorsten Naab$^3$, Micha\l~Hanasz$^4$, Stefanie Walch$^5$\\
$^1$Leibniz-Institut f\"{u}r Astrophysik Potsdam (AIP), An der Sternwarte 16, 14482 Potsdam, Germany\\
$^2$Heidelberger Institut f\"{u}r theoretische Studien, Schloss-Wolfsbrunnenweg 35, 69118 Heidelberg, Germany\\
$^3$Max-Planck-Institut f\"{u}r Astrophysik, Karl-Schwarzschild-Str. 1, 85741 Garching, Germany\\
$^4$Centre for Astronomy, Nicolaus Copernicus University, Faculty of Physics, Astronomy and Informatics, Grudziadzka 5, PL-87100 Toru\'n, Poland\\
$^5$Physikalisches Institut, Universit\"{a}t zu K\"{o}ln, Z\"{u}lpicher Str. 77, 50937 K\"{o}ln, Germany
}

\date{Accepted XXX. Received YYY; in original form ZZZ}

\pubyear{2018}

\begin{document}
\begin{NoHyper}
\label{firstpage}
\pagerange{\pageref{firstpage}--\pageref{lastpage}}
\maketitle

\begin{abstract}
We investigate the impact of cosmic rays (CRs) on galactic outflows from a multi-phase interstellar medium with solar neighbourhood conditions. The three-dimensional magneto-hydrodynamical simulations include CRs as a relativistic fluid in the advection-diffusion approximation. The thermal and chemical state of the ISM is computed with a non-equilibrium chemical network. We find that CRs (injected with 10 \% of the supernova energy) efficiently support the launching of outflows and strongly affect their phase structure. Outflows leaving the midplane are denser ($\rho \sim 10^{-26}\,\mathrm{g\,cm}^{-3}$), colder ($\sim 10^4\,\mathrm{K}$), and slower ($\sim 30\,\mathrm{km\,s}^{-1}$) if CRs are considered in addition to thermal SNe. The CR supported outflows are also smoother, in particular at larger heights ($> 1\,\mathrm{kpc}$ above the midplane) without the direct impact of SN explosions. Approximately $5\% - 25\%$ of the injected CR energy is lost via hadronic cooling. Smaller diffusion coefficients lead to slightly larger hadronic losses but allow for steeper CR pressure gradients, stronger outflows and larger accelerations. Up to a height of $z \sim1\,\mathrm{kpc}$ there are large volumes in approximate pressure equilibrium between thermal and CR component. At larger altitudes the CR pressure is $10-100$ times as large as the thermal counterpart. More than $\sim 1\,\mathrm{kpc}$ away from the midplane, CRs provide the dominant gas acceleration mechanism.
\end{abstract}


%%%%%%%%%%%%%%%%%%%%%%%%%%%%%%%%%%%%%%%%%%%%%%%%%%

%%%%%%%%%%%%%%%%% BODY OF PAPER %%%%%%%%%%%%%%%%%%

%\tableofcontents
%\newpage

\section{Introduction}

Galactic outflows are widely observed in star-forming galaxies \citep{VeilleuxCecilBlandHawthorn2005, HeckmanThompson2017}. They are important for the cycle of gas and with that the distribution of metals within the galaxy and the circumgalactic medium. Some fraction of the outflowing gas moves at velocities greater than the escape velocity and leaves the galaxy. The remaining gas forms a fountain flow, which eventually falls back onto the galaxy and influences subsequent star formation \citep[see][for a review of the theoretical models]{SomervilleDave2015, NaabOstriker2017}.

Outflows are powered by stellar feedback processes and are launched from the multiphase interstellar medium (ISM). As a consequence, the outflows themselves are multiphase in nature. Observational studies report outflows in a wide range of temperatures and chemical phases. Ionised gas in outflows has been found by \citet{PettiniEtAl2001, ShapleyEtAl2003, SteidelEtAl2010, ErbEtAl2012, NewmanEtAl2012b, GenzelEtAl2014, HeckmanEtAl2015, ChisholmEtAl2017}. Neutral outflowing gas is reported in studies by e.g. \citet{HeckmanEtAl2000, Martin2005, RupkeVeilleuxSanders2005, ChenEtAl2010, ContursiEtAl2013}. Even molecular gas is detected in outflows of e.g. M82 \citep{WeissEtAl1999, LeroyEtAl2015, ChisholmMatsushita2016} and other galaxies. Mass loading factors from 0.01 to 10 are observed \citep{VeilleuxCecilBlandHawthorn2005, NewmanEtAl2012b, GenzelEtAl2014, HeckmanThompson2017}, where the mass loading factor is defined as the ratio of outflow rate to star formation rate. For neutral and molecular gas the outflow rates decrease as a function of distance from the midplane \citep{ChisholmMatsushita2016}. Simulations of the supernova (SN)-driven ISM confirm the multiphase nature of outflows in a variety of galactic environments \citep{SILCC1, GirichidisSILCC2, LiBryanOstriker2017, KimOstriker2018}.

Stellar feedback influences the ambient medium by processes like radiation, stellar winds and SNe \citep{RogersPittard2013, HaidEtAl2016, GeenEtAl2015, GattoEtAl2015, PetersEtAl2017a, RahnerEtAl2017}. Important for the launching of outflows are the collective feedback effects, e.g. via the formation of superbubbles \citep{MacLowMcCray1988, WuenschEtAl2008}. Amoung the feedback processes, SNe are considered one of the most efficient drivers of interstellar turbulence and galactic fountain flows \citep{deAvillezBreitschwerdt2004, deAvillezBreitschwerdt2005, JoungMacLowBryan2009, KimKimOstriker2011, HillEtAl2012, KimOstrikerKim2013, GattoEtAl2015, SILCC1, GirichidisSILCC2, PadoanEtAl2016, GattoEtAl2017}. On galactic scales they might be important for the powering of galactic winds \citep{AumerEtAl2014, HopkinsEtAl2014, MarinacciPakmorSpringel2014, SchayeEtAl2015, HuEtAl2016}. However, there is also evidence that SNe are not sufficient to explain the turbulent motions in the ISM \citep{IbanezMejiaEtAl2016, IbanezMejiaEtAl2017, SeifriedEtAl2018}. On the other hand, in extreme environments outflows might also be triggered by turbulence and thermal runaway without energy input from SNe \citep{SurScannapiecoOstriker2016}.

Apart from the thermal and kinetic energy release of SNe, the strong shocks of SN remnants are efficient accelerators of cosmic rays (CRs) via diffusive shock acceleration \citep{AxfordEtAl1977, Bell1978, BlandfordOstriker1978, Schlickeiser1989}. Approximately $10\%$ of the fiducial SN energy of $10^{51}\,\mathrm{erg}$ can be converted into the non-thermal CR component \citep{HelderEtAl2012, MorlinoCaprioli2012, AckermannEtAl2013}. Most of the energy is deposited in protons with a momentum of a few GeV/$c$ \citep[see reviews by][]{StrongMoskalenkoPtuskin2007, GrenierBlackStrong2015}. Due to frequent scattering off magnetic field irregularities the CRs can be treated as a relativistic fluid with significantly different properties compared to the thermal counterpart \citep[see reviews by][]{Zweibel2013, Zweibel2017}. By not cooling efficiently CRs provide a long-lived energy reservoir. In addition, they can diffuse or stream relative to the gas motions, which allows the CR energy to be deposited independent of the gas velocities. The fiducial galactic estimates for the diffusion coefficients are large, so CRs with energies around a few GeV typically diffuse throughout the ISM on time scales much shorter than the dynamical time or the turbulent turnover time. The CR pressure gradients that develop over scales of kiloparsecs create an additional driving mechanism for galactic outflows. Detailed one-dimensional models find steady outflow solutions with wind velocities exceeding the escape velocity of the galaxy \citep{BreitschwerdtMcKenzieVoelk1991, DorfiBreitschwerdt2012, BustardZweibelDOnghia2016, BustardZweibelCotter2017, MaoOstriker2018} and are in agreement with observational constraints \citep{EverettSchillerZweibel2010}. Also three-dimensional numerical models support the picture of CR-driven outflows on galactic scales \citep{UhligEtAl2012, HanaszEtAl2013, BoothEtAl2013, SalemBryan2014, PakmorEtAl2016, RuszkowskiYangZweibel2017, JacobEtAl2018} as well as on ISM scales \citep{GirichidisEtAl2016CR, SimpsonEtAl2016, FarberEtAl2018}. The importance of CRs for the driving of outflows is probably overestimated in most of theses studies due to the insufficient and unresolved treatment of other feedback mechanisms, mostly SN feedback.

CRs in a simplified advection-diffusion approximation have proven to efficiently support the launching of outflows. However, the detailed interplay between the thermal and CR component of stellar feedback is still not understood. The efficiency of launching winds from the galaxy is strongly influenced by the details of the gas distribution in the disc and the spatial and temporal correlation of the main driving mechanisms \citep[see][for a review]{NaabOstriker2017}. Accurately modelling the dynamics and thermal properties of the ISM together with the most energetic accelerators for outflows is essential to understand the origin of fast winds with high mass loading. The aim of this study is to model the SN-driven interstellar medium including dynamically coupled CR feedback. Using a chemical network to account for appropriate cooling of the gas as well as (self-)shielding we investigate the relative importance of thermally driven compared to CR supported outflows.

\section{Numerical Methods and simulation parameters}

\begin{table*}
\begin{minipage}{0.9\textwidth}
  \begin{center}
  \caption{List of simulations and their main parameters}
  \label{tab:list-of-simulations}
  \begin{tabular}{lcccccccc}
    \hline
    name & $E_\mathrm{therm}$ & $E_\mathrm{CR}$ & $\mathsf{K}_\parallel$ & $\mathsf{K}_\perp$ & self-   & sim. time & min. cell size & comment\\
         & (erg)             & (erg)             & (cm$^2$\,s$^{-1}$)    & (cm$^2$\,s$^{-1}$) & gravity &  (Myr) & (pc) & \\
    \hline
    \texttt{noCR}      & $10^{51}$ & $0$       & $\phantom{3\,\times\,}10^{28}$  & $\phantom{3\,\times\,}10^{26}$  & no  & 150 &\phantom{1}3.9 & \\
    \texttt{CR-smlK} & $10^{51}$ & $10^{50}$ & $\phantom{3\,\times\,}10^{28}$  & $\phantom{3\,\times\,}10^{26}$  & no  & 150 &\phantom{1}3.9 & \\
    \texttt{CR-smlK-peakSN} & $10^{51}$ & $10^{50}$ & $\phantom{3\,\times\,}10^{28}$  & $\phantom{3\,\times\,}10^{26}$  & no  & 150 &\phantom{1}3.9 & SNe in density peaks\\    
    \texttt{CR-medK} & $10^{51}$ & $10^{50}$ & $3\times10^{28}$                & $3\times10^{26}$                & no  & 150 &\phantom{1}3.9 & \\
    \texttt{CR-medK-loc}$\zeta_\mathrm{CR}$ & $10^{51}$ & $10^{50}$ & $3\times10^{28}$                & $3\times10^{26}$                & no  & 100 & \phantom{1}3.9 & local $\zeta_\mathrm{CR}$\\
    \hline
    \texttt{noCR-sg}      & $10^{51}$ & $0$       & $\phantom{3\,\times\,}10^{28}$  & $\phantom{3\,\times\,}10^{26}$  & yes  & 100 &\phantom{1}3.9 & \\
    \texttt{CR-smlK-sg} & $10^{51}$ & $10^{50}$ & $\phantom{3\,\times\,}10^{28}$  & $\phantom{3\,\times\,}10^{26}$  & yes  & 100 &\phantom{1}3.9 & \\
    \texttt{CR-medK-sg} & $10^{51}$ & $10^{50}$ & $3\times10^{28}$                & $3\times10^{26}$                & yes  & 100 &\phantom{1}3.9 & \\
    \texttt{CR-medK-loc}$\zeta_\mathrm{CR}$\texttt{-sg} & $10^{51}$ & $10^{50}$ & $3\times10^{28}$                & $3\times10^{26}$                & yes  & 100 &\phantom{1}3.9 & local $\zeta_\mathrm{CR}$\\
  \hline
    \texttt{noCR-lo}      & $10^{51}$ & $0$       & $\phantom{3\,\times\,}10^{28}$  & $\phantom{3\,\times\,}10^{26}$  & no  & 150 &15.6 & \\
    \texttt{CR-tinyK-lo} & $10^{51}$ & $10^{50}$ & $\phantom{3\,\times\,}10^{27}$  & $\phantom{3\,\times\,}10^{25}$  & no  & 150 &15.6 & \\
    \texttt{CR-smlK-lo} & $10^{51}$ & $10^{50}$ & $\phantom{3\,\times\,}10^{28}$  & $\phantom{3\,\times\,}10^{26}$  & no  & 150 &15.6 & \\
    \texttt{CR-smlK-peakSN-lo} & $10^{51}$ & $10^{50}$ & $\phantom{3\,\times\,}10^{28}$  & $\phantom{3\,\times\,}10^{26}$  & no  & 150 &15.6 & SNe in density peaks\\
    \texttt{CR-medK-lo} & $10^{51}$ & $10^{50}$ & $3\times10^{28}$                & $3\times10^{26}$                & no  & 150 &15.6 & \\
    \texttt{CR-lrgK-lo} & $10^{51}$ & $10^{50}$ & $\phantom{3,\times\,}10^{29}$  & $\phantom{3\,\times\,}10^{27}$  & no  & 150 &15.6 & \\
  \hline  
  \end{tabular}
  \end{center}
  
  \medskip
  We group the runs in three different sets: the main runs without self-gravity, the runs including self-gravity and the low-resolution runs. From left to right we list the name, the thermal ($E_\mathrm{therm}$) and CR energy ($E_\mathrm{CR}$) injected per SN, the parallel ($\mathsf{K}_\parallel$) and perpendicular ($\mathsf{K}_\perp$) diffusion coefficient, whether self-gravity is included, the simulation time, and the resolution at the highest level of refinement. The last column adds comments where appropriate.
\end{minipage}
\end{table*}

Both the simulation setup as well as most of the used physical modules are described in detail in \citet{SILCC1} and \citet{GirichidisSILCC2}. Here, we only present the modules in short, except for the cosmic ray solver, which is presented in more detail.

We use the hydrodynamical code \textsc{FLASH} in version 4 (\citealt{FLASH00,DubeyEtAl2008}, \url{http://flash.uchicago.edu/site/}) which is parallelised using the Message Passing Interface (MPI).
The Eulerian grid divides the computational domain in blocks of $8^3$ cells which can be adaptively refined (Adaptive Mesh Refinement, AMR). The magneto-hydrodynamic (MHD) equations are solved using the modified HLLR3 finite-volume scheme for ideal MHD \citep{Bouchut2007, Bouchut2010, Waagan2009, Waagan2011}. The directionally split solver ensures positivity of the density and pressure by construction and is suitable for high-Mach-number flows. The combined system of equations that we solve numerically is
\begin{align}
  \frac{\partial\rho}{\partial t} + \nabla\cdot\left(\rho\mathbf{v}\right) &= 0\\
  \frac{\partial\rho\mathbf{v}}{\partial t} + \nabla\cdot\left(\rho\mathbf{v}\mathbf{v}^\mathrm{T} - \frac{\mathbf{B}\mathbf{B}^\mathrm{T}}{4\pi}\right) + \nabla P_\mathrm{tot} &= \rho\mathbf{g}+\dot{\mathbf{q}}_{_\mathrm{SN}}\\
  \frac{\partial e}{\partial t} + \nabla\cdot\left[\left(e + P_\mathrm{tot}\right)\mathbf{v} - \frac{\mathbf{B}(\mathbf{B}\cdot\mathbf{v})}{4\pi}\right] &= \notag\\
  \rho\mathbf{v}\cdot\mathbf{g} + \nabla\cdot(\mathsf{K}\nabla e_{_\mathrm{CR}}) + \dot{u}_\mathrm{chem} + \dot{u}_{_\mathrm{SN}} + Q_{_\mathrm{CR}}&\\
  \frac{\partial\mathbf{B}}{\partial t} - \nabla \times \left(\mathbf{v}\times\mathbf{B}\right) &= 0\\
  \frac{\partial e_{_\mathrm{CR}}}{\partial t} + \nabla\cdot(e_{_\mathrm{CR}}\mathbf{v}) &= \notag\\
  -P_{_\mathrm{CR}}\nabla\cdot\mathbf{v} + \nabla\cdot(\mathsf{K}\nabla e_{_\mathrm{CR}}) + Q_{_\mathrm{CR}},
\end{align}
Here, $\rho$ is the mass density, $\mathbf{v}$ is the velocity, and $\mathbf{B}$ is the magnetic field. The energetic changes due to chemical transitions, radiative cooling as well as background UV and X-ray heating are captured in $\dot{u}_\mathrm{chem}$ and are explained in more detail below. The injection of thermal energy ($\dot{u}_{_\mathrm{SN}}$) or momentum ($\dot{q}_{_\mathrm{SN}}$) by SNe is outlined in Sec.~\ref{sec:simulation-parameters}. The total energy density,
\begin{equation}
    e = \rho v^2/2 + e_\mathrm{th} + e_{_\mathrm{CR}} + B^2/8\pi,
\end{equation}
includes kinetic, thermal, CR, and magnetic contributions. We evolve the CR energy density, $e_{_\mathrm{CR}}$, separately. The total pressure is
\begin{alignat}{3}\label{eq:total-pressure}
  P_\mathrm{tot} &= P_\mathrm{th} &&+ \,P_{_\mathrm{CR}} &&+ \,P_\mathrm{mag}\\
  &= (\gamma-1)e_\mathrm{th} &&+ \,(\gamma_{_\mathrm{CR}}-1)e_{_\mathrm{CR}} &&+ \,B^2/8\pi.
\end{alignat}
The closure relation for the system, the equation of state, combines the different contributions from CR and thermal pressure in an effective adiabatic index, $\gamma_\mathrm{eff}$,
\begin{equation}
  \gamma_\mathrm{eff} = \frac{\gamma P_\mathrm{th} + \gamma_{_\mathrm{CR}} P_{_\mathrm{CR}}}{P_\mathrm{th} + P_{_\mathrm{CR}}},
\end{equation}
where we set $\gamma = 5/3$ and $\gamma_\mathrm{CR} = 4/3$ for gas and CRs, respectively. For the CR diffusion tensor, $\mathsf{K}$, we assume a value of $10^{28}\,\mathrm{cm^2\,s^{-1}}$ parallel and $10^{26}\,\mathrm{cm^2\,s^{-1}}$ perpendicular to the magnetic field lines \citep[e.g.,][]{StrongMoskalenkoPtuskin2007,NavaGabici2013}. The CR source term, $Q_{_\mathrm{CR}}$, includes SNe as CR sources ($Q_{_\mathrm{SN,CR}}$) and hadronic losses ($\Lambda_\mathrm{hadr}$),
\begin{align}
Q_{_\mathrm{CR}} = Q_{_\mathrm{SN,CR}} + \Lambda_\mathrm{hadr}.
\end{align}
For each SN we inject $10^{50}\,\mathrm{erg}$ in CRs, so $10\%$ of the fiducial SN energy. For the hadrononic losses we follow the prescription in \citet{PfrommerEtAl2017},
\begin{equation}
\label{eq:hadr-losses}
\Lambda_\mathrm{hadr} = -7.44\times10^{-16}\,\left(\frac{n_\mathrm{e}}{\mathrm{cm}^{-3}}\right)\,\left(\frac{e_{_\mathrm{CR}}}{\mathrm{erg\,cm}^{-3}}\right)\,\mathrm{erg~s}^{-1}~\mathrm{cm}^{-3}
\end{equation}
assuming a steady state spectrum. Here, $n_e$ is number density of free electrons.

For the gravitational acceleration we distinguish between the effects of self-gravity by solving the Poisson equation using the tree-based method described in \citet{WuenschEtAl2018} and an external potential which accounts for the acceleration of the stellar component of the disc as well as dark matter. We use tabulated values based on the Milky-Way potential by \citet{KuijkenGilmore1989}.

To follow the chemical state of the gas and compute radiative cooling accurately, we use a chemical network including ionised (H$^+$), atomic (H) and molecular hydrogen (H$_2$) as well as singly ionized carbon (C$^+$) and carbon monoxide (CO). This allows us to include non-equilibrium abundances as in \citet{GloverMacLow2007a} and \citet{MicicEtAl2012} using the carbon chemistry by \citet{NelsonLanger1997}. We assume a temporally constant UV interstellar radiation field of $G_0=1.7$ \citep{Habing1968, Draine1978, Draine2011}, where $G_0$ is the integrated energy density from $6-13.6\,\mathrm{eV}$ in units of the Habing estimate $u_\mathrm{Hab}=5.29\times10^{-14}\,\mathrm{erg\,cm}^{−3}$,
\begin{equation}
G_0\equiv\frac{u(6-13.6\,\mathrm{eV})}{u_\mathrm{Hab}} = \frac{u(6-13.6\,\mathrm{eV})}{5.29\times10^{-14}\,\mathrm{erg\,cm}^{−3}}.
\end{equation}
This radiation field is locally attenuated based on how strongly shielded the computational cell is. The column densities and the derived optical depth in every cell are computed using the TreeCol algorithm \citep{ClarkGloverKlessen2012}. The column density dependent attenuation factor follows \citet{GloverClark2012b}.

For radiative cooling we follow the atomic and molecular cooling functions of \citet{GloverEtAl2010} and \citet{GloverClark2012b}. High-temperature cooling above $10^4\,\mathrm{K}$ is based on the rates described in \citet{GnatFerland2012}. Heating includes a constant cosmic ray ionisation ($\zeta_\mathrm{CR}=3\times10^{-17}\,s^{-1}$) and corresponding heating rate \citep{GoldsmithLanger1978}, X-ray heating \citep{WolfireEtAl1995}, and photoelectric heating that is coupled to the optical depth \citep{BakesTielens1994, Bergin2004, WolfireEtAl2003}. We assume a constant dust-to-gas ratio of 0.01 using dust opacities of \citet{MathisMezgerPanagia1983} and \citet{OssenkopfHenning1994}.

\subsection{Simulation Parameters}
\label{sec:simulation-parameters}

We set up a stratified box with a size of $0.5\,\mathrm{kpc}\times0.5\,\mathrm{kpc}\times\pm10\,\mathrm{kpc}$. In $x$ and $y$ direction we use periodic boundary conditions. For the boundary in $z$ direction we allow gas to leave but not enter the box. For the CRs we only allow for negative gradients across the $z$ boundary, i.e. we only allow diffusion out of the box. The initial gas density is described by a Gaussian distribution in $z$ with a scale height of $30\,\mathrm{pc}$. There are no initial density perturbations in the $xy$ plane. We apply a lower density floor of $10^{-28}\,\mathrm{g~cm}^{-1}$ and set the temperature such that the gas is in pressure equilibrium. In the densest regions of the disc at $z=0$ this corresponds to a minimum temperature of $4600\,\mathrm{K}$ and purely atomic hydrogen. At the lowest density ($\rho_\mathrm{min}=10^{-28}\,\mathrm{g\,cm}^{-3}$) at large $|z|$ we set $T=4\times10^8\,\mathrm{K}$ and assume the gas to be fully ionised. The gas surface density is $10\,M_\odot\mathrm{pc}^{-2}$, which yields a total mass of $2.5\times10^6\,M_\odot$ in the box. Initially the gas is at rest.

The magnetic field is oriented along the $x$ direction with initial field strengths in the midplane of $B_{x,0}=1\,\mathrm{nG}$. The magnitude of $B$ scales with the square root of the density, $B_x(z)=B_{x,0}\,[\rho(z)/\rho(z=0)]^{1/2}$. We do not introduce a random component of the magnetic field. We chose a very low initial field to allow for a self-consistently generated field by means of local compression and small-scale dynamos.

Feedback from stars is included as SN heating at a constant SN rate based on the average surface density of the disc. We use the Kennicutt-Schmidt \citep{KennicuttSchmidt1998} relation to set the star formation rate, which is then converted into a SN rate using the stellar initial mass function of \citet{Chabrier2003}. Three different types of SNe are considered. We let 20\% of the SNe explode as type~Ia, which explode at random $x$ and $y$ positions and a Gaussian distribution in $z$ with a scale height of $300\,\mathrm{pc}$. The remaining 80\% are SNe of type~II with a scale height of $120\,\mathrm{pc}$. The latter component is split into $2/5$ of individual SNe (representing run-away OB stars with random positions in the $xy$-plane) and $3/5$ of SNe, which are associated with star clusters and the resulting clustered explosion of massive stars as SNe \citep{CowieSongailaYork1979}. This clustered component is expected to be an important agent in driving superbubbles \citep{MacLowMcCray1988}. The life time of the clusters is set to $40\,\mathrm{Myr}$. Each cluster contains $7-20$ SNe, randomly drawn from a powerlaw distribution, where the probability is $P\propto N^{-2}$ with $N$ being the number of SNe. All clusters, their associated SNe and the explosion times are determined beforehand and stored in a table. Concerning the positioning of the SNe we again distinguish between the types of explosions. For SNe of type~Ia and the random component of type~II, we draw random positions a priory according to the vertical distribution. The clusters are placed in density maxima. When the first SN of a cluster is due according to the total list of SNe, we find the densest region in the simulation box and place the cluster at this position. This mimics the fact that stellar clusters are born in dense molecular clouds. For the rest of the clusters life time it remains at this fixed position. For the SNe this means that the first few explosions occur in dense regions until a hot bubble is created in which the subsequent SNe of the cluster explode. 
For each explosion we inject $10^{51}\,\mathrm{erg}$ of thermal energy. If we resolve the Sedov-Taylor radius with at least 4 cells we inject thermal energy into a spherical volume that encompasses a mass of $800\,M_\odot$, i.e. $\dot{u}_{\mathrm{SN}}dt=10^{51}\,\mathrm{erg}$ where $dt$ is the simulation time step and $\dot{\mathbf{q}}_{_\mathrm{SN}}=\mathbf{0}$. If this resolution requirement is not fulfilled we set $\dot{u}_{\mathrm{SN}}$ such that the gas in the injection region with a radius of 4 cells is at $T=10^4\,\mathrm{K}$. The injection of mechanical feedback ($\dot{\mathbf{q}}_{_\mathrm{SN}}$) follows \citet{BlondinEtAl1998}, see also \citet{GattoEtAl2015}. The radially outward pointing velocities are computed as $v_\mathrm{inj}=p_\mathrm{ST}/M_\mathrm{inj}$, where $p_\mathrm{ST}=2.5\times10^5\,(n/\mathrm{cm}^{-3})^{-2/17}\,M_\odot\,\mathrm{km\,s}^{-1}$. Here, $M_\mathrm{inj}$ and $n$ are the total mass and average number density of the injection region. The advantage of this somewhat artificial setup for SN feedback is that the time distribution as well as the ratio between individual and clustered SNe is the same for all runs. By setting the clusters in the densest region of the box when the first SN of a cluster explodes, we also ensure similar efficiencies for the clustered SN distribution. This allows for a selfconsistent investigation of the CR feedback. 

Table~\ref{tab:list-of-simulations} lists all simulations with their main parameters. The columns show (from left to right) the name of the simulation, the thermal energy per SN, the CR energy per SN, the parallel and perpendicular diffusion coefficient, the switch for self-gravity, the total simulation time, the cell size at the highest level of refinement, and additional comments where appropriate. The main five simulations are the first group in the table. We distinguish between a purely thermal run \texttt{noCR} with only $10^{51}\,\mathrm{erg}$ per SN and CR runs where we inject $10^{50}\,\mathrm{erg}$ per SN in addition to the thermal energy. We \emph{add} this additional energy because we would like to ensure the same thermal energy input for simplicity. Reducing the thermal input by 10\% does not change the result. Among the CR runs we vary the diffusion coefficients (\texttt{CR-smlK}, \texttt{CR-medK}) and adapt the local CR ionisation rate to the CR energy field (\texttt{CR-medK-loc}$\zeta_\mathrm{CR}$), where we scale $\zeta_\mathrm{CR}$ linearly with $e_{_\mathrm{CR}}$,
\begin{align}
\zeta_\mathrm{CR} = 3\times10^{-17}\,\mathrm{s}^{-1} \, \left(\frac{e_{_\mathrm{CR}}}{10^{-12}\,\mathrm{erg\,cm}^{-3}}\right).
\end{align}
In addition we change the positioning of the SNe from the clustered driving described above to a run, where we use \emph{peak SN driving} (\texttt{CR-smlK-peakSN}), which means that we place every SN in the densest region of the simulation box. In order to have a partially structured gas distribution we start run \texttt{CR-smlK-peakSN} with random SN positions for the first $\sim10\mathrm{Myr}$ before switching to the peak driving mode. The second group of simulations includes self-gravity, which is problematic in our setups with the given resolution and the limited feedback processes, see Sec.~\ref{sec:discussion} and Appendix~\ref{sec:app-self-gravity}. The third group of simulations are low-resolution runs that allow us to explore a larger variation of the diffusion coefficient as well as convergence tests for the thermal and CR related effects (Sec.~\ref{sec:app-resolution}). 

\section{Morphological evolution}

\subsection{Vertical evolution}

\begin{figure*}
\begin{minipage}{\textwidth}
\includegraphics[height=0.45\textheight]{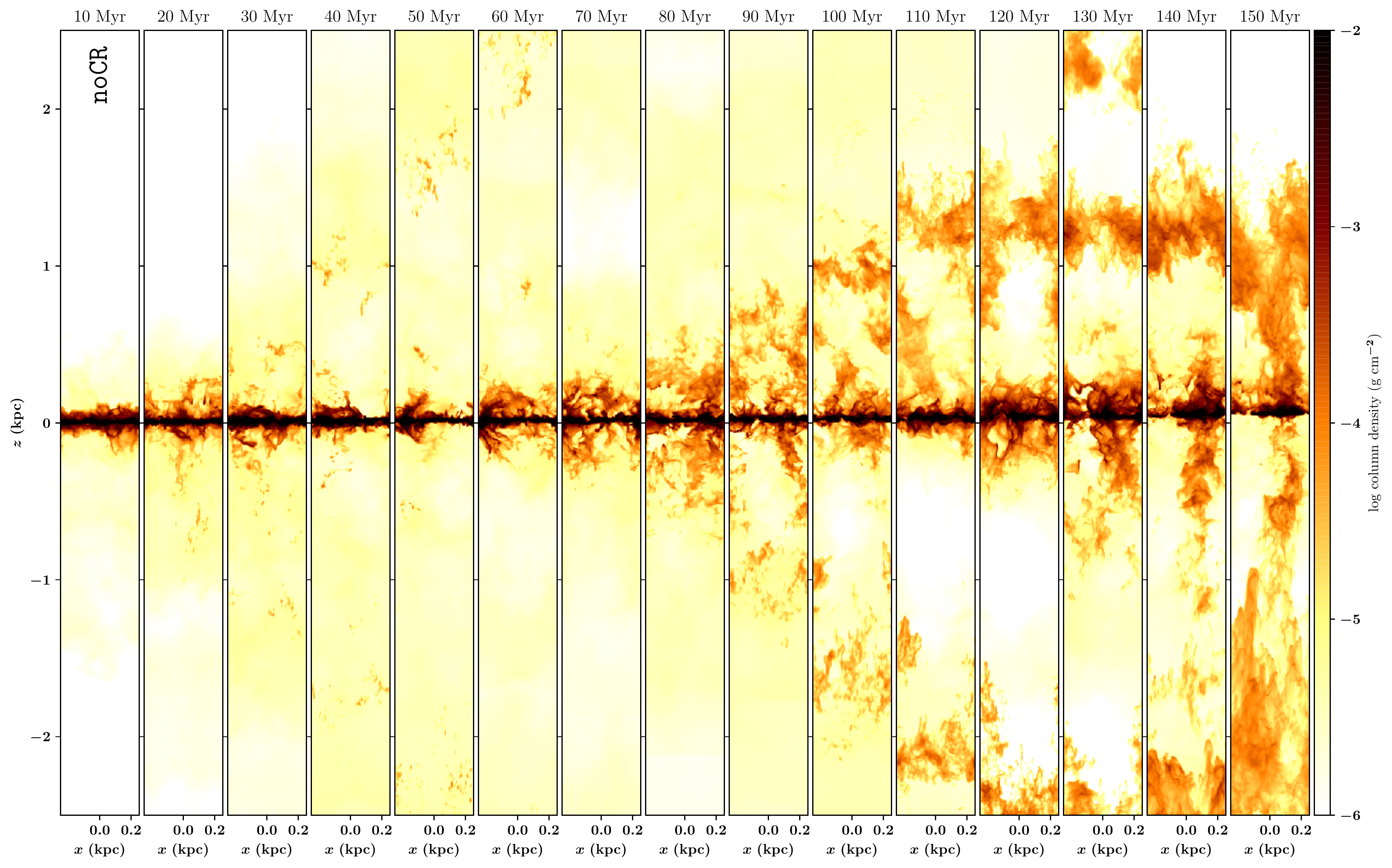}\\
\includegraphics[height=0.45\textheight]{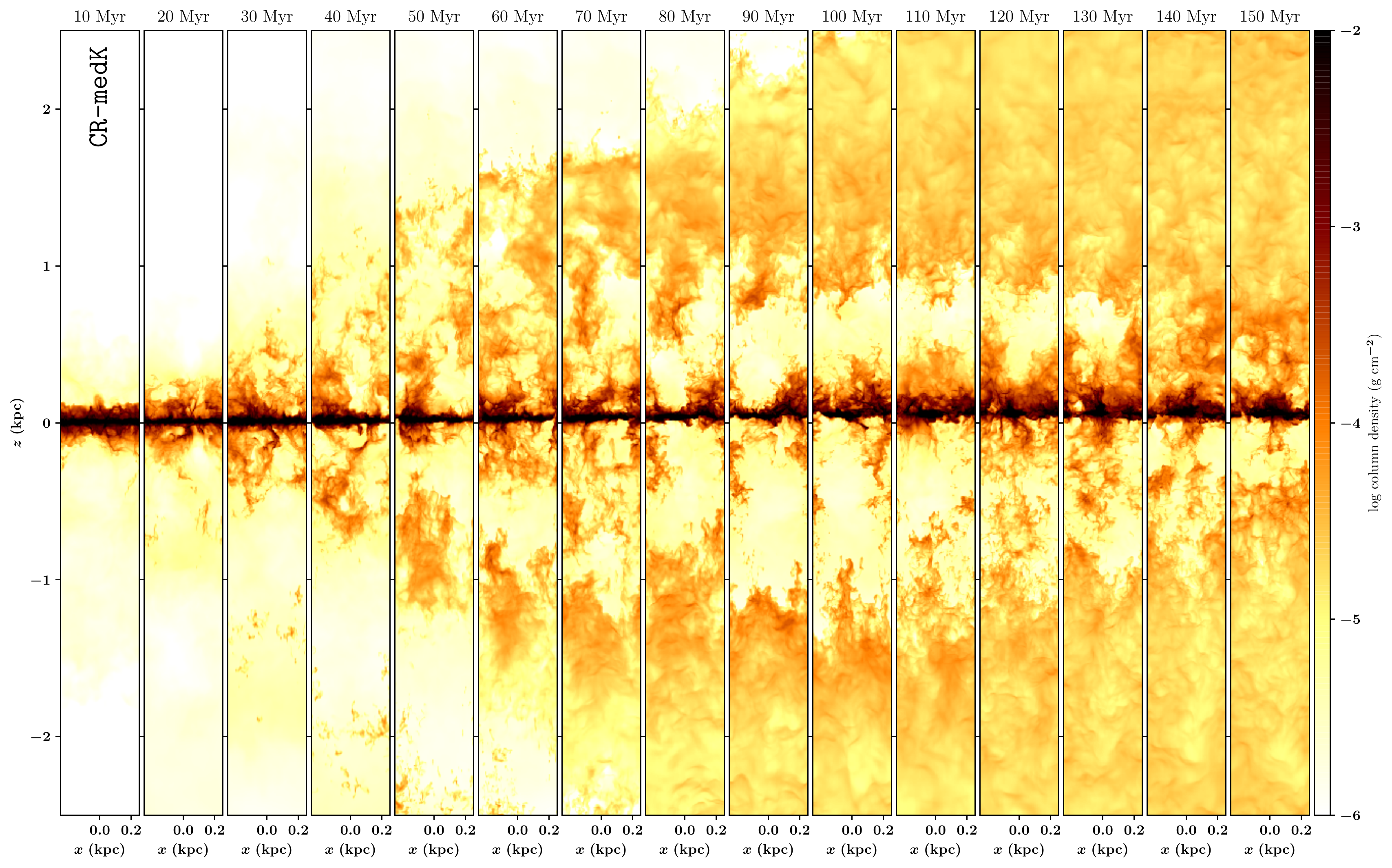}\\
\end{minipage}
\caption{Time evolution of the edge-on column density for simulations \texttt{noCR} (top) and \texttt{CR-medK} (bottom) in steps of $10\,\mathrm{Myr}$. The purely thermal run (top panel) creates patchy outflows with large voids in the halo. Towards the end of the simulation some gas begins to fall back towards the disc. The CR run launches outflows that are highly structured up to a height of $1\,\mathrm{kpc}$, which is the threshold with SN injection. Above that height the gas redistributes with low column-density contrasts but at significantly higher column densities than in run \texttt{noCR}.}
\label{fig:Hill-plots-00-10s-ng}
\end{figure*}

\begin{figure*}
\begin{minipage}{\textwidth}
\includegraphics[height=0.45\textheight]{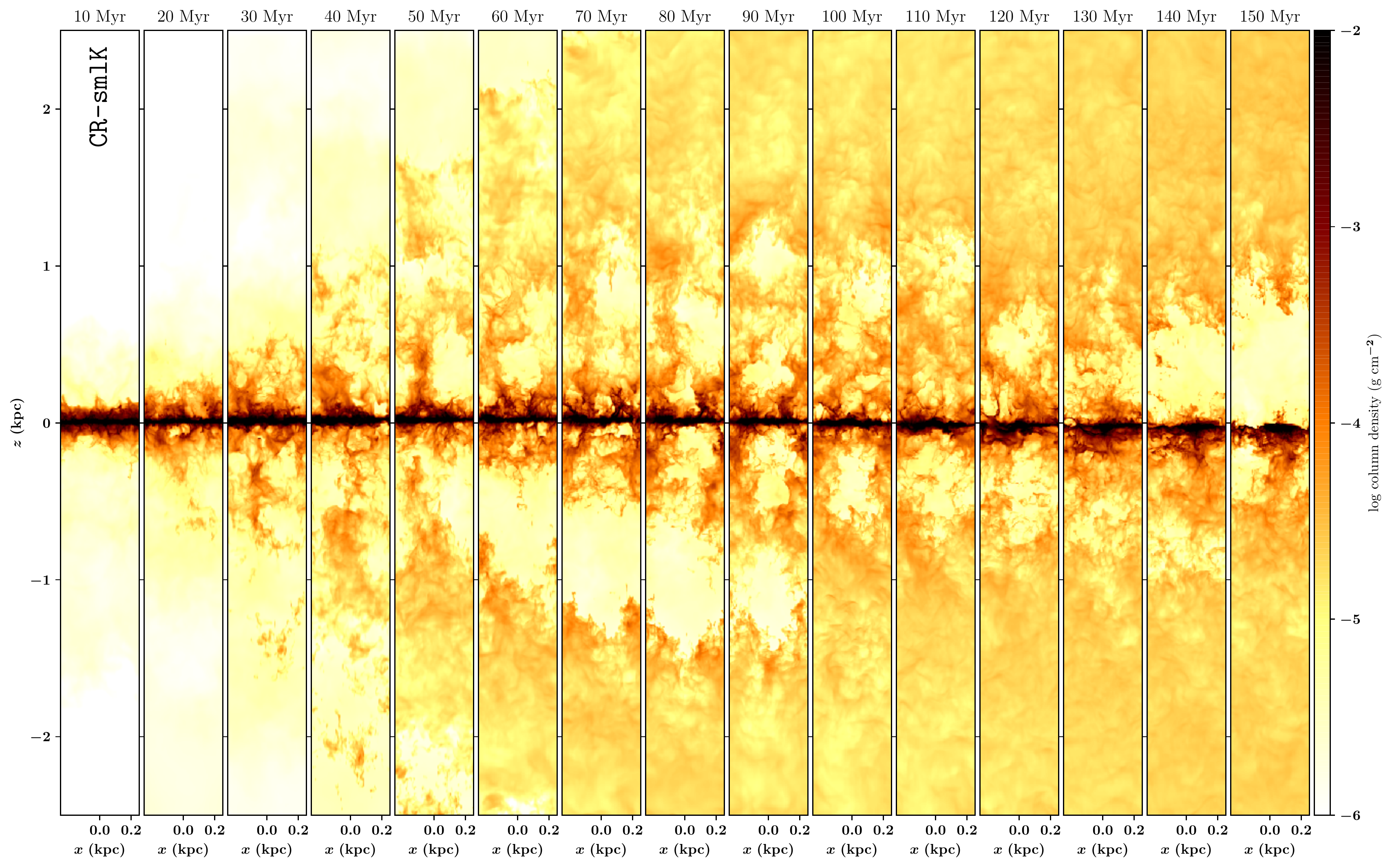}
\includegraphics[height=0.45\textheight]{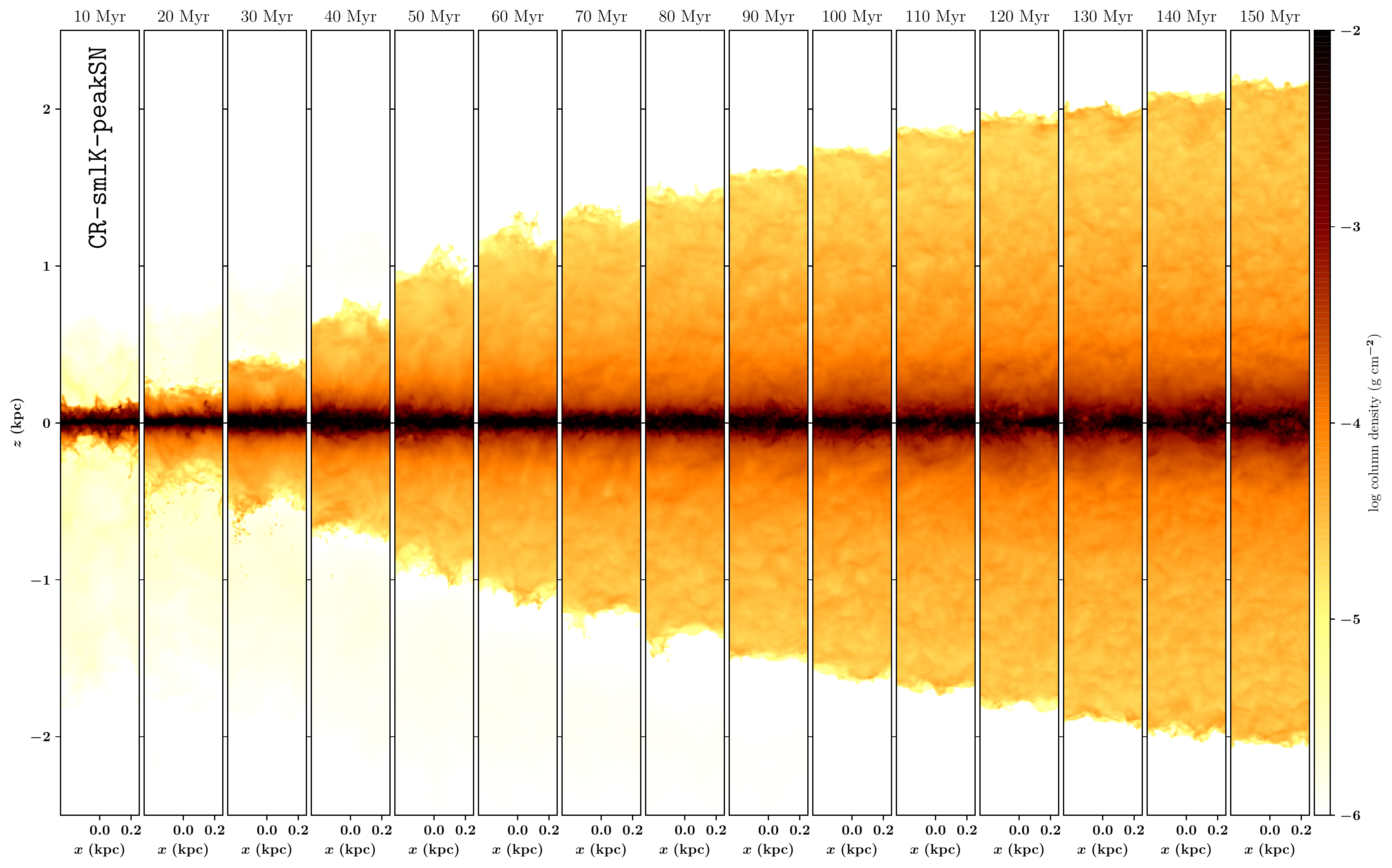}
\end{minipage}
\caption{Time evolution of the edge-on column density for simualtions \texttt{CR-smlK} (top) and \texttt{CR-smlK-peakSN} (bottom) in steps of $10\,\mathrm{Myr}$. Run \texttt{CR-smlK} develops a structured outflow up to a height of $1\,\mathrm{kpc}$ and a smooth gas distribution above that altitude, similar to \texttt{CR-medK}. In simulation \texttt{CR-smlK-peakSN} all SNe explode in the dense region of the disc and their thermal impact for driving outflows is insignificant \citep[see][]{GirichidisSILCC2}. The mainly CR-driven outflow fills the halo on longer time scales compared to \texttt{CR-smlK}. The halo gas is dense with a very smooth distribution.}
\label{fig:Hill-plots-10m-10sp-ng}
\end{figure*}

In all runs with clustered SN positions the feedback creates dense cold filaments and hot voids in the disc after a few Myr. The simulation with peak driving efficiently destroys the local overdensities by placing the SNe in the density peaks, which results in mostly diffuse warm gas without hot voids. The vertical evolution of the simulations is shown in Fig.~\ref{fig:Hill-plots-00-10s-ng} for simulations \texttt{noCR} and \texttt{CR-medK} and in Fig.~\ref{fig:Hill-plots-10m-10sp-ng} for simulations \texttt{CR-smlK} and \texttt{CR-smlK-peakSN}. Shown are column density projections in steps of $10\,\mathrm{Myr}$ for the entire evolution time. Shortly after the beginning of the simulations the SNe start driving gas out of the plane into the halo. In the purely thermal run the outflowing gas is at low density and the outflows have a patchy structure. In the simulations including CRs the gas that starts filling the halo is initially also highly structured but overall denser. The gas filling the halo in the simulation with peak driving is very smooth and dense. At later times the differences between the driving mechanisms become more evident. In simulation \texttt{noCR} a significant fraction of the halo gas is at low densities with high-density patches. In simulations \texttt{CR-medK} and \texttt{CR-smlK} the gas up to a hight of $\sim1.5\,\mathrm{kpc}$ remains structured throughout the simulation time. The gas above that altitude forms a perceptibly smoother distribution. In the case of peak driving the outflows are very smooth from the very beginning and noticeably slower compared to the CR runs with clustered SNe. Whereas the lifted gas in run \texttt{CR-smlK} reaches a height of $2\,\mathrm{kpc}$ after $70-80\,\mathrm{Myr}$, it takes $130-140\,\mathrm{Myr}$ in simulation \texttt{CR-smlK-peakSN}.

Variations of the diffusion coefficient by half an order of magnitude do not change the overall appearance (Fig.~\ref{fig:Hill-plots-00-10s-ng} bottom and Fig.~\ref{fig:Hill-plots-10m-10sp-ng} top). Similarly, a locally varying CR ionisation rate coupled to the CR energy density (\texttt{CR-medK-loc}$\zeta_\mathrm{CR}$, not shown here) results in minor differences compared to the corresponding run with constant rate (\texttt{CR-medK}).

We note that the midplane of the disc is temporally offset from $z=0$. These vertical motions are due to clustered SNe that are located slightly above or below the midplane. As the stellar clusters (and their associated SNe) have life times of $40\,\mathrm{Myr}$, the collective dynamical effect lasts for a noticeable fraction of the simulation time.  

\subsection{Dense gas and chemical composition in the disc}

\begin{figure}
  \centering
  \includegraphics[width=8cm]{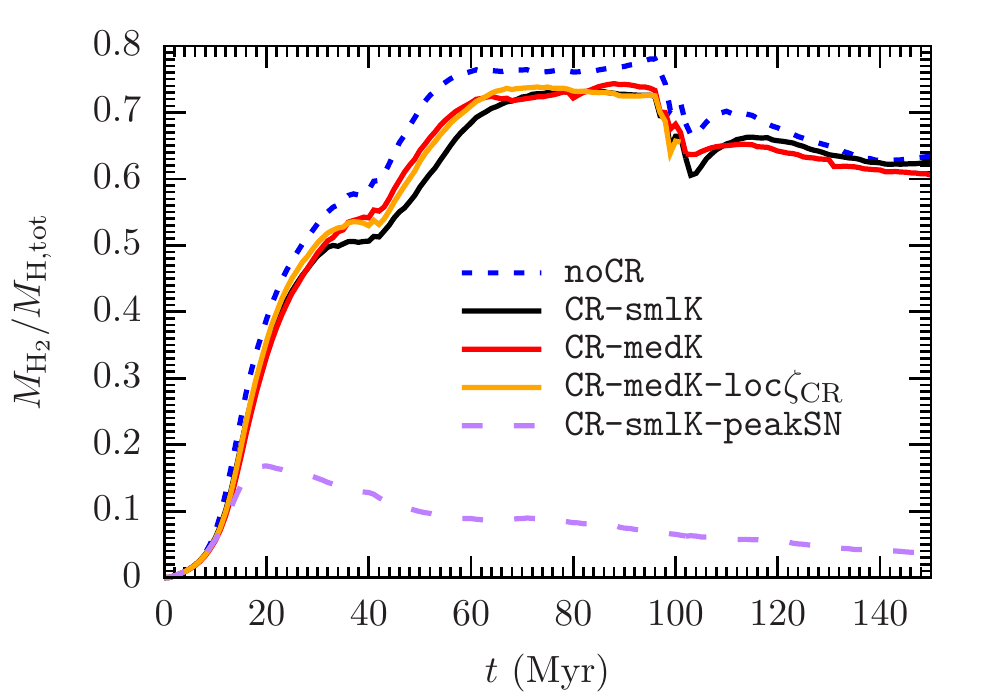}
  \caption{The time evolution of the mass fraction of molecular hydrogen ($M_\mathrm{H_2}/M_\mathrm{H_2,tot}$) is relatively similar for all runs with clustered SN driving and saturates at around $0.6-0.7$. The thermal run forms slightly more H$_2$. At $t=40\,\mathrm{Myr}$ and $t=100\,\mathrm{Myr}$ a few new massive clusters are activated, which stalls the formation and reduces the total fraction of H$_2$, respectively. The SNe in density peaks in \texttt{CR-smlK-peakSN} destroy all overdensities and reduce the formation of H$_2$.}
  \label{fig:mass-fraction-H2-time-evol}
\end{figure}

The lack of self-gravity does not allow for a detailed analysis of the structure and the chemical composition in the disc \citep[see][]{SILCC1}, so we only use the molecular hydrogen as a tracer for dense regions and the amount of gas in dense structures. In Fig.~\ref{fig:mass-fraction-H2-time-evol} we plot the global H$_2$ mass fraction in the simulation box ($M_\mathrm{H_2}/M_\mathrm{H,tot}$) over time. For all simulations with clustered SNe the fraction of molecular gas varies over time between $\sim60-80\%$ excluding the initial phase of H$_2$ formation. We find slightly lower H$_2$ fractions for the CR runs compared to the run without CRs. This is consistent with the simulations by \citet{SimpsonEtAl2016}, where the inclusion of CRs reduce the amount of dense gas and the resulting star formation rate. However, we use a different driving and clustering of the feedback, so we cannot draw a quantitative conclusion here. At around $40\,\mathrm{Myr}$ and $100\,\mathrm{Myr}$ a few new clusters of SNe are activated, i.e. the densest positions in the simulation box are found and the cluster with all subsequent SNe is fixed to this position. This retards the formation of molecular gas at the former point in time and reduces the fraction of H$_2$ at the latter time. The run with SNe in density peaks only forms a negligible amount of molecular gas, which has also been reported in \citet{SILCC1} and \citet{GirichidisSILCC2} without CRs but including self-gravity.

We note that our former studies \citep{SILCC1, GirichidisSILCC2} did not find molecular gas if self-gravity is not included. The two differences in the present study that influence the results are the broader vertical distribution of SN as well as the different external potential. The first SILCC study used a vertical scale height of only $50\,\mathrm{pc}$ for the type~II SNe, which deposits noticeably more energy in the disc region compared to the distribution with a scale height of $120\,\mathrm{pc}$ used in this study. The external gravitational potential based on \citet{DickeyLockman1990} is stronger by a factor of 2-3 (depending on height) compared to the isothermal sheet approximation used in former work. The combined effect allows the gas to form molecular hydrogen in the disc.

\subsection{Magnetic field strength}

\begin{figure}
\includegraphics[width=8cm]{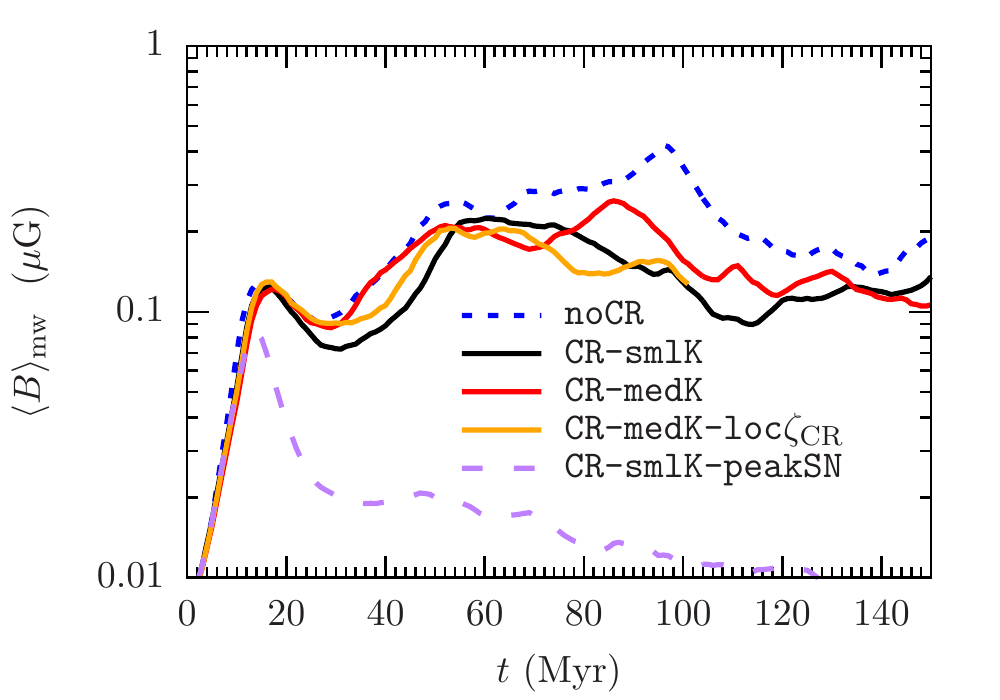}
\caption{Mass weighted magnetic field strength over time, $\langle B\rangle_\mathrm{mw}$. The initially weak field is amplified to about $0.1-0.3\,\mu\mathrm{G}$ after $\sim20\,\mathrm{Myr}$. In simulation \texttt{CR-smlK-peakSN} the field is about one order of magnitude weaker than in the other runs.}
\label{fig:magnetic-field-strength-time-evol}
\end{figure}

We start with a very small magnetic field of $1\,\mathrm{nG}$ in order to self-consistently follow the dynamical evolution and magnetic field amplification. Fig.~\ref{fig:magnetic-field-strength-time-evol} shows the time evolution of the magnetic field strength. We compute the mass weighted values,
\begin{equation}
\langle B\rangle_\mathrm{mw}=M_\mathrm{tot}^{-1}\sum_i m_i|\mathbf{B}_i|,
\end{equation}
with $m_i$ being the mass in cell $i$, $|\mathbf{B}_i|$ the corresponding magnetic field strength, and $M_\mathrm{tot}$ the total mass in the simulation box. The initial values are quickly amplified to the saturation value of $\langle B\rangle_\mathrm{mw}\sim0.1-0.3\,\mu\mathrm{G}$ after approximately $20\,\mathrm{Myr}$ with little difference between most of the simulations. We note that in all runs the magnetic field is lower than observed field strengths of a few $\mu\mathrm{G}$. In run \texttt{CR-smlK-peakSN} the driving of the gas by peak SNe is not strong enough to maintain the initial peak of the magnetic field strength and values decrease perceptibly to values that are an order of magnitude lower than in the other simulations. The field strengths in our simulations are clearly below the observed ones of a few $\mu\mathrm{G}$ in the galactic plane \cite[e.g.][]{Beck2009}. We mainly attribute this to the insufficient resolution for driving a small-scale dynamo and the missing shear due to differential rotation, but see also the discussion in Sec.~\ref{sec:discussion}. Finally, we are also missing the large scale magnetic field of the galaxy. In all cases the field is dominated by small-scale structures. We refrain from doing a detailed analysis of the field geometry here but illustrate the morphology below when describing the vertical structure.

\section{Vertical structure and vertical motions}

\begin{figure}
   \includegraphics[width=8cm]{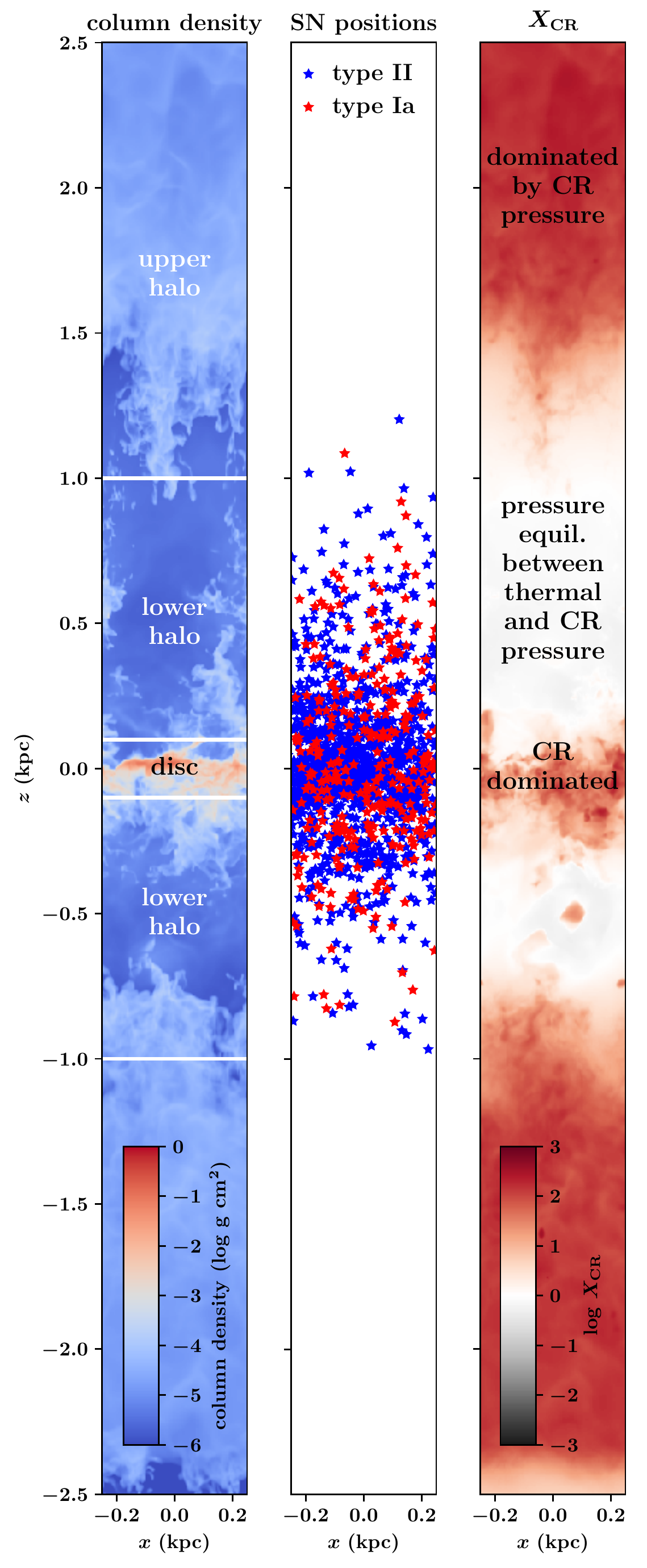}
   \caption{Sketch of the vertical sections for the detailed analysis presenting the column density (left) and the ratio of CR to thermal pressure, $X_\mathrm{CR}$ (right) at a time of $t=100\,\mathrm{Myr}$. The middle panel shows the positioning of all SNe up to $t=100\,\mathrm{Myr}$. We distinguish between three different volume for parts of the analysis: the disc ($|z|\le0.1\,\mathrm{kpc}$), the lower halo (between $0.1$ and $1\,\mathrm{kpc}$) and the upper halo (region between $1$ and $2.5\,\mathrm{kpc}$).}
   \label{fig:sketch-vertical-volumes}
\end{figure}

\begin{figure*}
\begin{minipage}{\textwidth}
\centering
\includegraphics[width=0.95\textwidth]{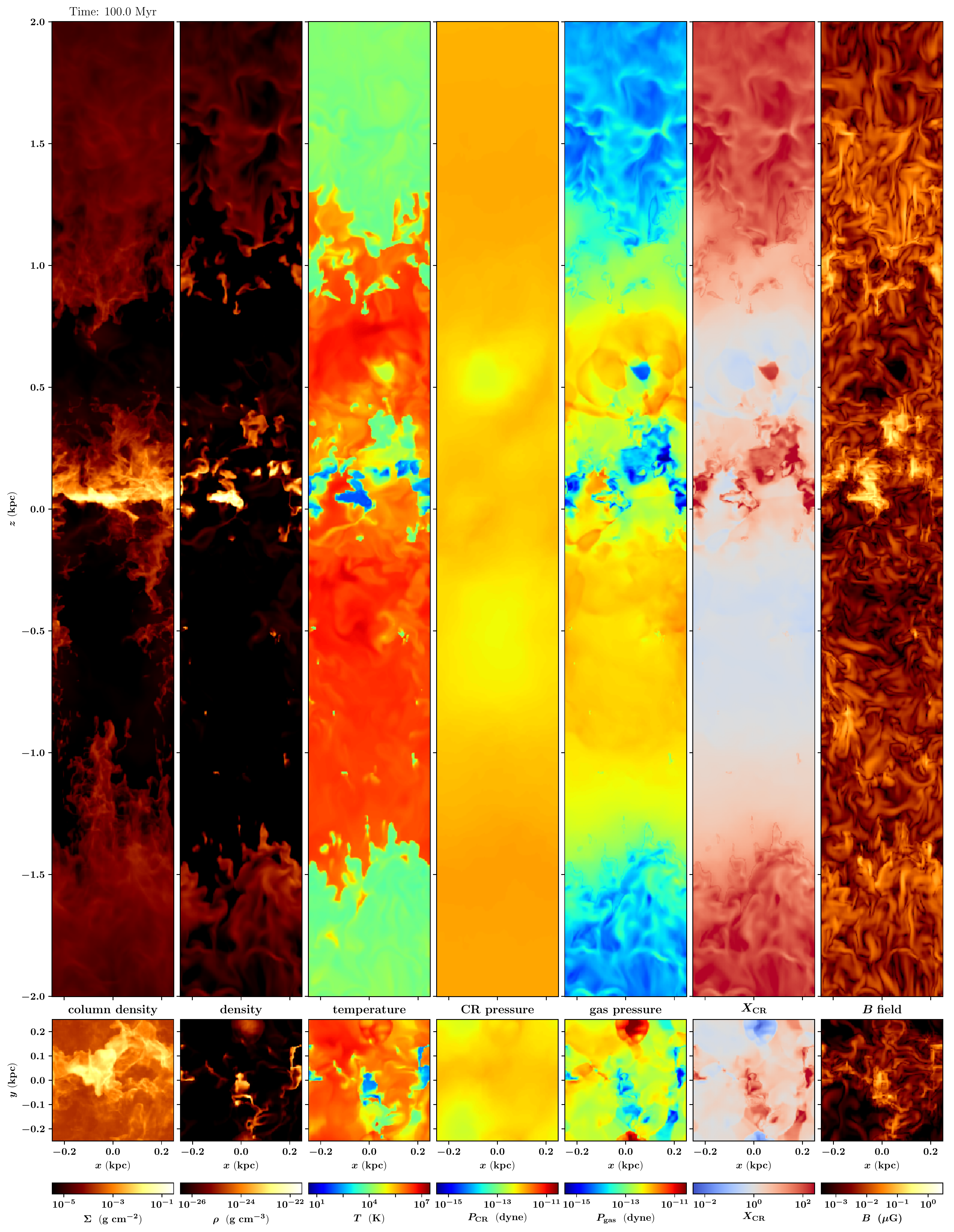}
\end{minipage}
\caption{Vertical structure of simulation \texttt{CR-medK} at $t=100\,\mathrm{Myr}$. The left panel shows the column density, the other panels present infinitesimal slices through the centre of the box with the density, the temperature, the CR pressure, the thermal pressure as well as the ratio of CR to thermal pressure. The CR simulations drive gas to heights above $1\,\mathrm{kpc}$, which is warm ($T\sim10^4\,\mathrm{K}$) and CR dominated. The CR pressure varies by only one order of magnitude, the thermal pressure by four orders of magnitude, which explains the strong variations in $X_\mathrm{CR}$. The magnetic field is tangled and reaches peak values of approximately $1\,\mu\mathrm{G}$ in the densest regions.}
\label{fig:vertical-profiles-overview}
\end{figure*}

\begin{figure*}
\begin{minipage}{\textwidth}
\includegraphics[width=\textwidth]{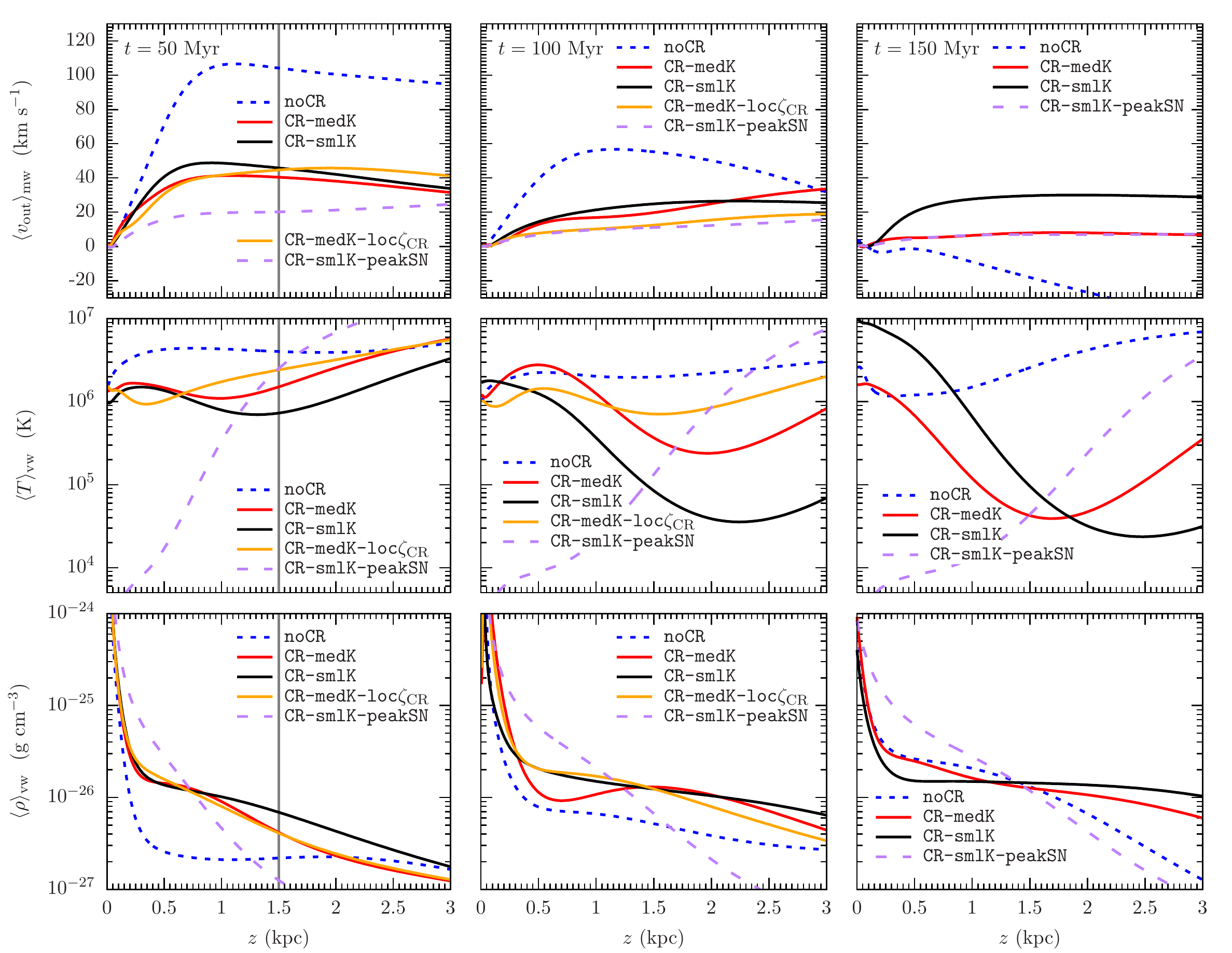}
\end{minipage}
\caption{Vertical profiles of the outward pointing velocity, the temperature, and the gas density (from top to bottom) at $50\,\mathrm{Myr}$ (left), $100\,\mathrm{Myr}$ (middle), and $150\,\mathrm{Myr}$ (right). The grey line in the left panels indicates the approximate height up to which outflowing gas has moved within the first $50\,\mathrm{Myr}$. Thermally driven outflows are twice as fast in the first $100\,\mathrm{Myr}$ and expand into a significantly hotter halo of lower density. After $t=100\,\mathrm{Myr}$ run \texttt{noCR} develops a fountain flow with infalling motions. CR supported outflows are an order of magnitude denser and lead to a colder halo. At $t=150\,\mathrm{Myr}$ the flow of the gas in \texttt{CR-medK} almost stalls. Only simulation \texttt{CR-smlK} continues to lift gas into the halo at a speed of $30\,\mathrm{km\,s}^{-1}$. \texttt{CR-smlK-peakSN} drives a slow, dense and warm outflow from a relatively cold disc compared to all runs with clustered SNe at random positions.}
\label{fig:vertical-profiles}
\end{figure*}

\begin{figure*}
\begin{minipage}{\textwidth}
\includegraphics[width=\textwidth]{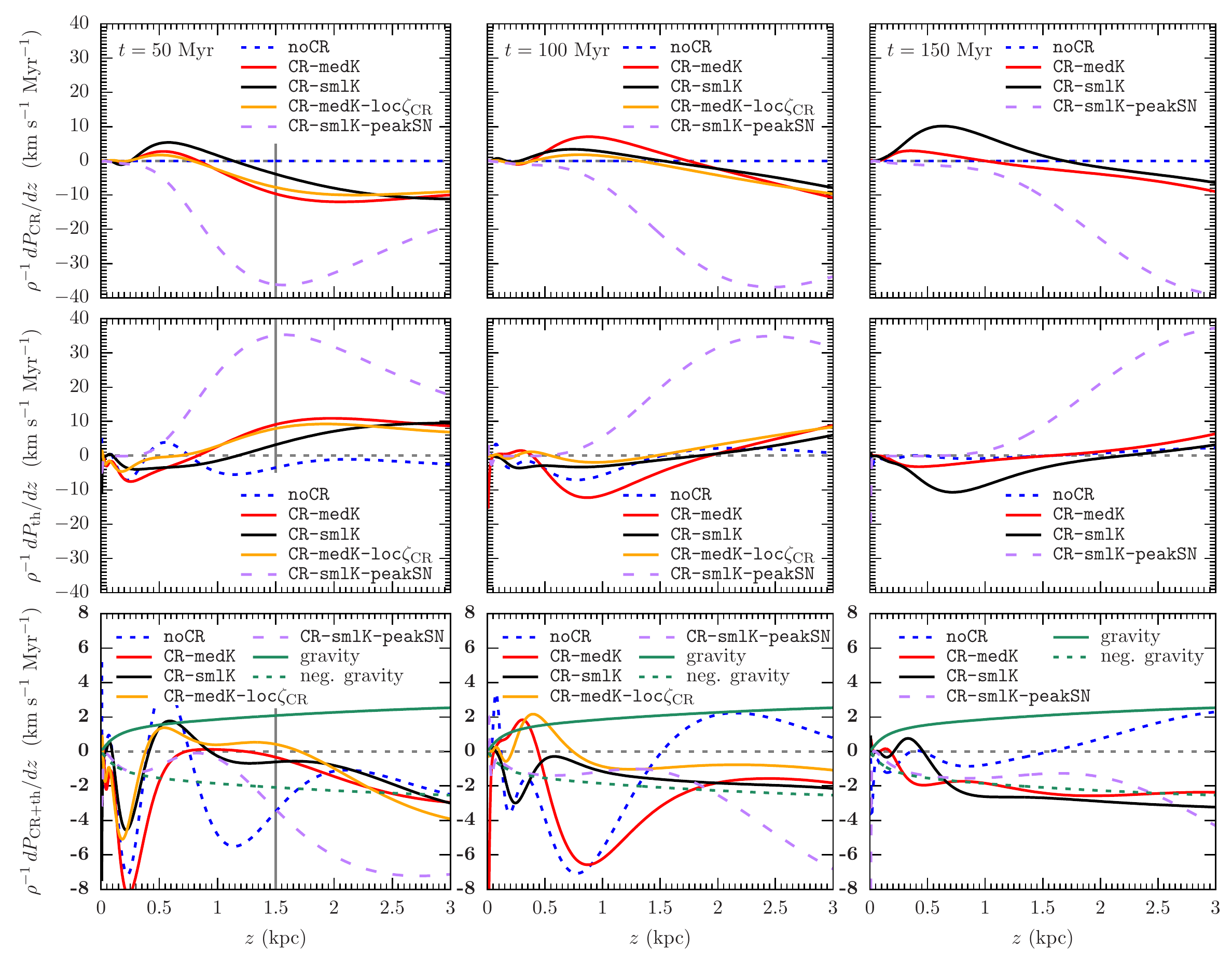}
\end{minipage}
\caption{Vertical profiles of the acceleration due to thermal and CR pressure. The upper and middle panels show CR and thermal contribution. The bottom panel combines both pressures as well as the \emph{negative} gravitational acceleration to better compare it to the outflow generating force. Negative numbers indicate an acceleration pointing away from the midplane. Thermal and CR profiles show overall opposing vertical trends and partially cancel (note smaller range in ordinate in bottom panels). The net acceleration in the thermal runs shows stronger variations both along the vertical axes as well as in time. The CR profiles are smoother and more steady. After $100\,\mathrm{Myr}$ the acceleration in the CR runs almost reaches the strength of the gravitational attraction, after $150\,\mathrm{Myr}$ the pressures in simulation \texttt{CR-smlK} are about 1.5 as strong as the gravitational attraction.}
\label{fig:vertical-profiles-force}
\end{figure*}

%\subsection{Vertical zones}

For our further discussion we focus on three vertical zones of the computational box, which are related to different dominating physical processes. Fig.~\ref{fig:sketch-vertical-volumes} illustrates these zones. The left panel shows the projected gas density of one of the CR simulations at $t=100\,\mathrm{Myr}$. The middle panel shows the positions of all SNe up to $t=100\,\mathrm{Myr}$. In the right panel we plot the ratio of thermal to CR pressure, averaged along the $y$-direction. The \emph{disc} is considered to be the volume with $|z|\le100\,\mathrm{pc}$. The region between $0.1<|z|/\mathrm{kpc}\le1$ is called \emph{lower halo} in the following. The cut at $1\,\mathrm{kpc}$ marks the limit of the direct SN injection regions. The \emph{upper halo} marks the region between $1$ and $2.5\,\mathrm{kpc}$ height above the midplane. For our analysis we limit the height to a maximum of $2.5\,\mathrm{kpc}$ because most of the volume a larger altitudes is not affected by outflowing gas over the time scale of $150\,\mathrm{Myr}$.

\subsection{Vertical gas distribution}

An overview of the vertical distribution is depicted in Fig.~\ref{fig:vertical-profiles-overview} for simulation \texttt{CR-medK} at $100\,\mathrm{Myr}$. The top panels are the edge-on views of the box, the lower panels are the corresponding face-on plots. The left panel shows the integrated density, all other panels represent infinitesimal slices through the center of the box. Shown are the density, the gas temperature, the CR pressure, the gas pressure, the ratio of CR to gas pressure as well as the magnetic field strength. The disc region is patchy with cold clouds ($\sim10\,\mathrm{K}$) and hot voids ($10^6-10^7\,\mathrm{K}$). The CR pressure is relatively smooth with an average value of approximately $10^{-12}\,\mathrm{erg\,cm}^{-3}$. The thermal pressure varies by more than four orders of magnitude, which explains the local variations of the ratio of CR to gas pressure $X_\mathrm{CR}$ in the right panel. We note that the disc region as well as the upper halo are dominated by CR pressure, whereas the lower halo region is in approximate pressure equilibrium or slightly dominated by thermal pressure. The magnetic field intensity in the right panel indicates that the field is strongly tangled with structures reflecting the turbulent gas motions. This is expected since the magnetic field is dynamically unimportant. There is a noticeable but not very prominent correlation of the field strength with the density.

A peculiar feature in Fig.~\ref{fig:vertical-profiles-overview} is the cold spherical region at a height of $z\approx0.6\,\mathrm{kpc}$, which is a short-lived cold regions of a previously expanded SN remnant. The strong expansion results in a low temperature, a low gas pressure and a weak magnetic field. The dilution of CR energy is compensated by fast diffusion back into the low-temperature region. As a result the small variations in CR energy density results in a large value of $X_\mathrm{CR}$.

The vertical structure of the simulation box averaged over $x$ and $y$ is presented in Fig.~\ref{fig:vertical-profiles}. In the top panels we plot the mass weighted velocity oriented away from the disc, $\langle v_\mathrm{out}\rangle_\mathrm{mw}=M^{-1}\sum_i \mathrm{sgn}(z) m_i v_{z,i}$, where $i$ is the index set of all cells that intersect with the vertical position $\pm z$, $M$ is the total mass of that cells, and $m_i$ and $v_{z,i}$ are the individual cell masses and $z$ components of the velocities. The second and third row of panels show the volume weighted averages of the temperature, $\langle T\rangle_\mathrm{vw}=V^{-1}\sum_i V_i T_i$, and the total gas density, $\langle\rho\rangle_\mathrm{vw}=V^{-1}\sum_i V_i \rho_i$, with the total volume $V$ at $\pm z$ and the individual cell quantities $V_i$, $T_i$, and $\rho_i$ for volume, temperature and density, respectively. From left to right the panels show $t=50\,\mathrm{Myr}$, $t=100\,\mathrm{Myr}$, and $t=150\,\mathrm{Myr}$, respectively. At $t=50\,\mathrm{Myr}$ the outflowing gas has barely reached heights of $1.5\,\mathrm{kpc}$ (indicated by the gray vertical line), so the halo region above that height is still dominated by the pristine environment. The velocity profiles reveal that the thermally driven vertical motions are faster than the CR supported outflows at $50\,\mathrm{Myr}$ and $100\,\mathrm{Myr}$ by a factor of a few. There is little difference between the CR runs. However, after $t=150\,\mathrm{Myr}$ the velocities in \texttt{noCR} become negative indicating the fountain flow behaviour, in which some gas falls back towards the disc. In simulation \texttt{CR-medK} the velocities also decrease over time to result in an almost static gas distribution at the end. Only in the run with a small CR diffusion coefficient the gas is still moving outwards at speeds of $30\,\mathrm{km\,s}^{-1}$. The run with SNe in density peaks launches the slowest fountain flows that also almost stalls at the end of the simulations with almost indistinguishable velocity profiles from \texttt{CR-medK}. The temperature and density profiles are inversely correlated. This behaviour is strongest in simulation \texttt{CR-smlK-peakSN}, where the disc and lower halo are cold ($T<10^4\,\mathrm{K}$) and dense throughout the runtime. The advancing outflow manifests in a moving transition of the cold front over time. For \texttt{noCR} at $50\,\mathrm{Myr}$ the densities are an order of magnitude lower compared to the lower halos in \texttt{CR-smlK}, \texttt{CR-medK} and \texttt{CR-medK-loc}$\zeta_\mathrm{CR}$. The temperature is a factor of a few hotter. At $t=100\,\mathrm{Myr}$ the region above $1\,\mathrm{kpc}$ remains hot in the thermal run whereas the temperatures decrease to a few $10^5\,\mathrm{K}$ for \texttt{CR-medK} and few $10^4\,\mathrm{K}$ for \texttt{CR-smlK}. This strong temperature differences at large altitudes even slightly increase at $t=150\,\mathrm{Myr}$ with almost two orders of magnitude hotter gas in the thermal run. The very high temperatures in simulation \texttt{CR-smlK} for $|z|\lesssim500\,\mathrm{pc}$ at $150\,\mathrm{Myr}$ is a temporal feature of very low densities in the lower halo, which is nicely illustrated in the time evolution plot (Fig.~\ref{fig:Hill-plots-00-10s-ng}). The comparably fast changes in the density and velocity profiles of the thermal run suggest that CRs enlarge the time scales for fountain flows. 

\subsection{Vertical acceleration}

Fig.~\ref{fig:vertical-profiles-force} presents vertical acceleration profiles at $50\,\mathrm{Myr}$ (left), $100\,\mathrm{Myr}$ (middle), and $150\,\mathrm{Myr}$ (right). We use again a volume weighting along $x$ and $y$. In the left panels for $50\,\mathrm{Myr}$ we again indicate the approximate height that the outflowing gas has reached by a vertical line. The top panel depicts the contribution due to CRs, the middle one is for the thermal pressure. The bottom panel shows the sum of thermal and CR acceleration and also includes the gravitational acceleration. Negative values indicate an acceleration pointing away from the midplane. Total acceleration profiles that lie below the curve of negative gravity indicate outward pointing accelerations that can potentially unbind the gas from the disc. We would like to stress the smaller range of the ordinate in the bottom panel for better illustration. This is because the accelerations due to CR and thermal pressure gradient partially cancel. At $50\,\mathrm{Myr}$ the CRs provide an outward acceleration above $z\sim1\,\mathrm{kpc}$. The corresponding thermal pressure provides an opposite acceleration pattern such that the net acceleration due to pressure approximately vanishes. The purely thermal run provides only an outwardly pointing acceleration, which exceeds the gravitational attraction, below $1.5\,\mathrm{kpc}$. At a height of $1\,\mathrm{kpc}$ the acceleration in \texttt{noCR} is more than twice as strong as the gravitational acceleration, which corresponds directly to the large outflow velocities discussed above. After $t=100\,\mathrm{Myr}$ the features in the profiles have shifted to larger altitudes. For the CR runs the $P_\mathrm{CR}$ profile is still approximately opposite to the thermal profile but the sum of both now provide a net acceleration away from the disc, which is almost as strong as the gravitational attraction. Contrary, simulation \texttt{noCR} developed an inwardly pointing pressure gradient which supports gravity and contributes to the deceleration of the gas. At the end of the simulation time the net acceleration in the thermal run is still pointing towards the midplane for heights above $1.5\,\mathrm{kpc}$. At that height the gas started falling back to the disc. For the CR runs the total acceleration counteracts the gravitational attraction over a large range of the simulation box. For \texttt{CR-smlK} the outward pointing forces even exceed the opposite gravitational force by $30\%$.

\subsection{Outflows and halo gas}

\begin{figure*}
  \centering
  \includegraphics[width=\textwidth]{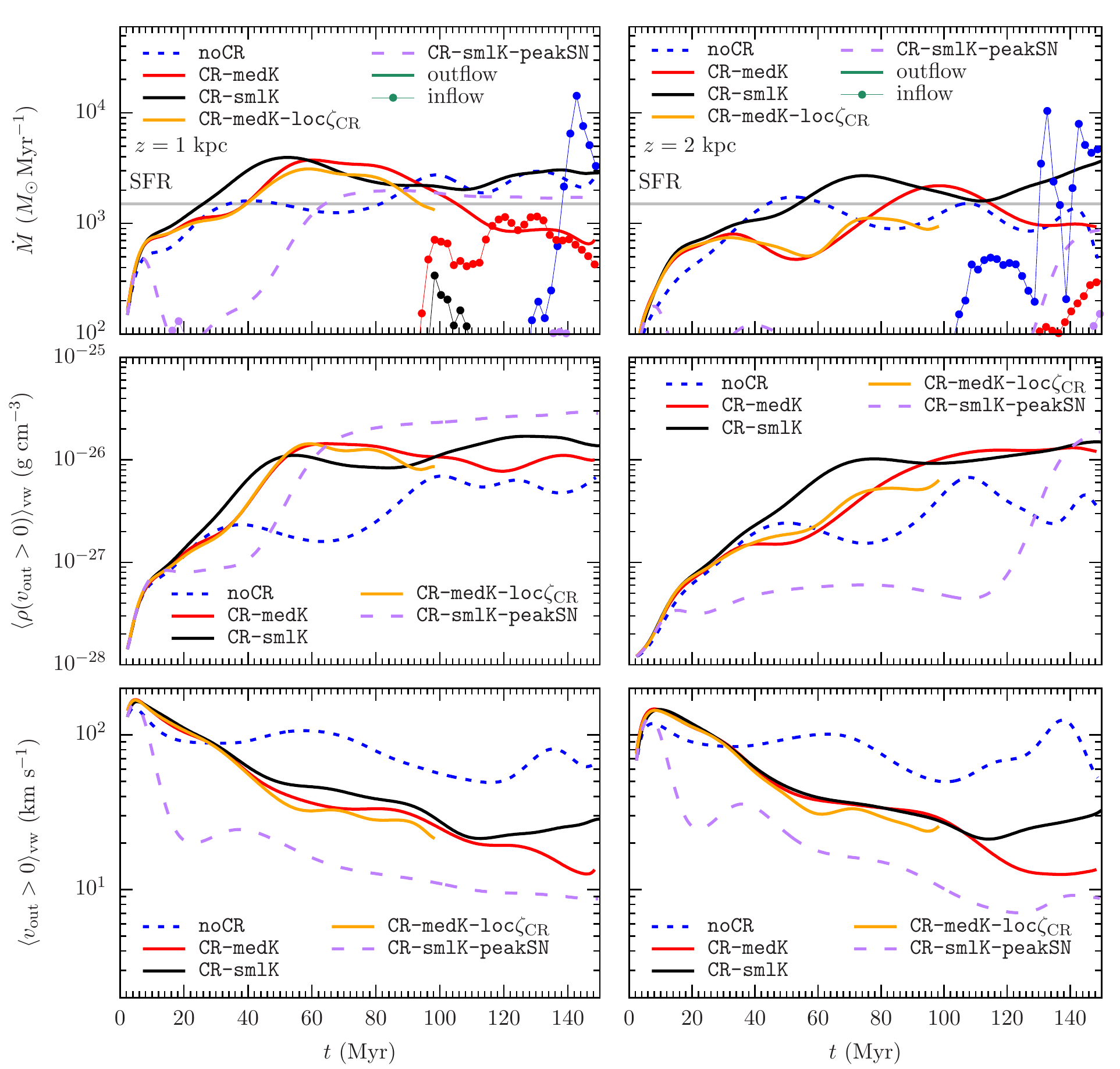}
  \caption{Outflow properties for all simulations over time at $1\,\mathrm{kpc}$ (left) and $2\,\mathrm{kpc}$ (right) height. From top to bottom we show the mass flux rate as well as volume average of the density and velocity of the outflowing gas at the measurement heights. In the top panels, the lines show the outflowing material and the points with thin lines the infall rate. The outflow rates vary over time around mean values of $1-2$ times the SFR with smaller average outflow rates for purely thermal SNe. After $110$ and $140\,\mathrm{Myr}$ the infall for \texttt{CR-medK} and \texttt{noCR} approach or exceed the outflow rate resulting in net infall at $|z|=1\,\mathrm{kpc}$, see also \texttt{noCR} for $t\gtrsim120\,\mathrm{Myr}$ at $|z|=2\,\mathrm{kpc}$. Independent of the measurement height the CR supported outflows are a factor of a few times denser and slower.}
  \label{fig:outflow-time-evol}
\end{figure*}

\begin{figure}
  \includegraphics[width=8cm]{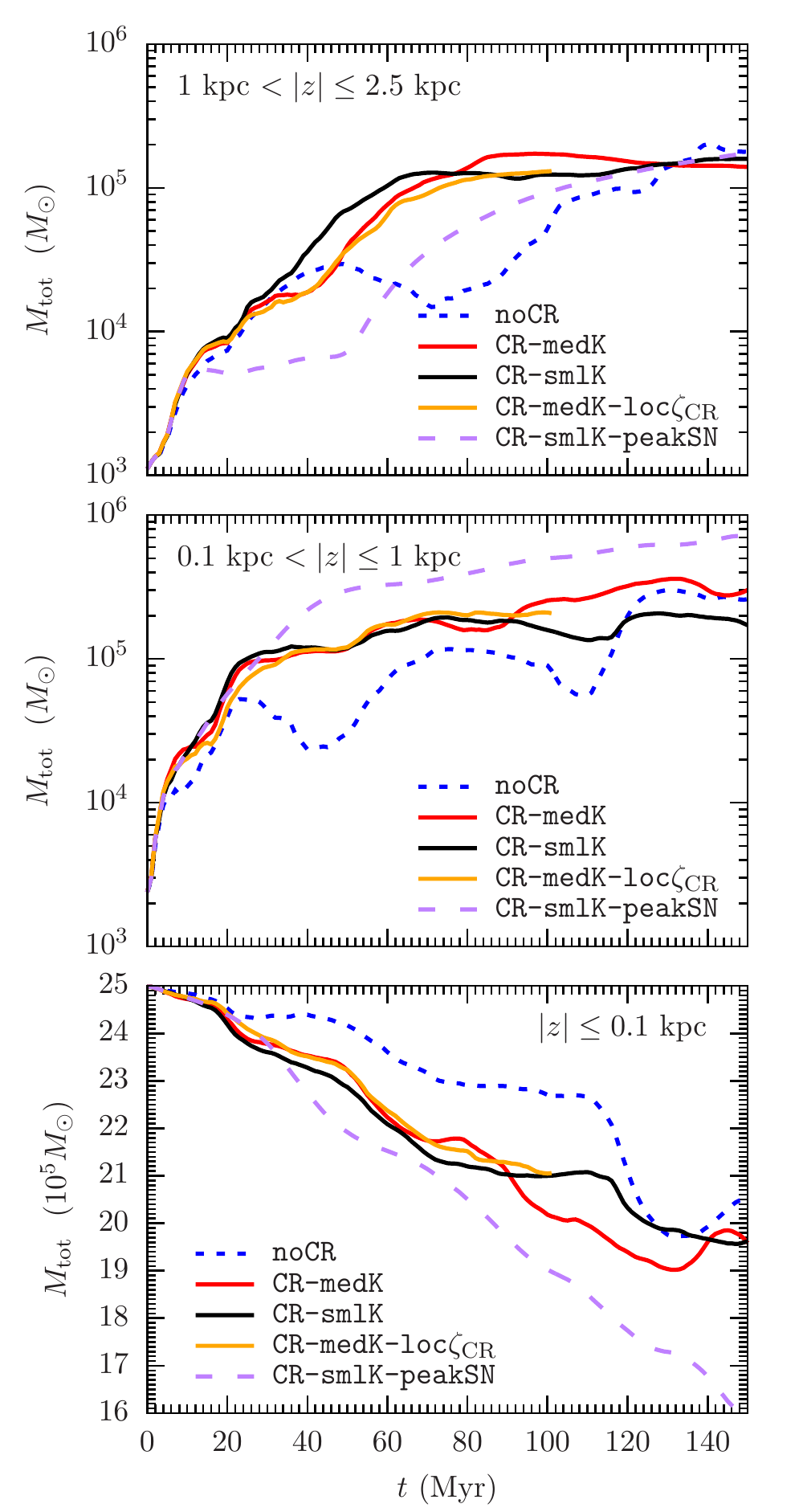}
  \caption{Gas mass in different volumes of the box. All setups generate outflows, so the disc ($|z|<0.1\,\mathrm{kpc}$) loses mass over time (note the linear scale). The CR supported outflows cause twice as much mass to leave the disc compared to the purely thermal run. This gas is lifted up to heights of $2\,\mathrm{kpc}$ with all CR runs depositing at least twice as much gas as run \texttt{noCR} for a significant fraction of the simulation time.}
  \label{fig:halo-mass-time-evol}
\end{figure}

\begin{figure}
   \centering
   \includegraphics[width=8cm]{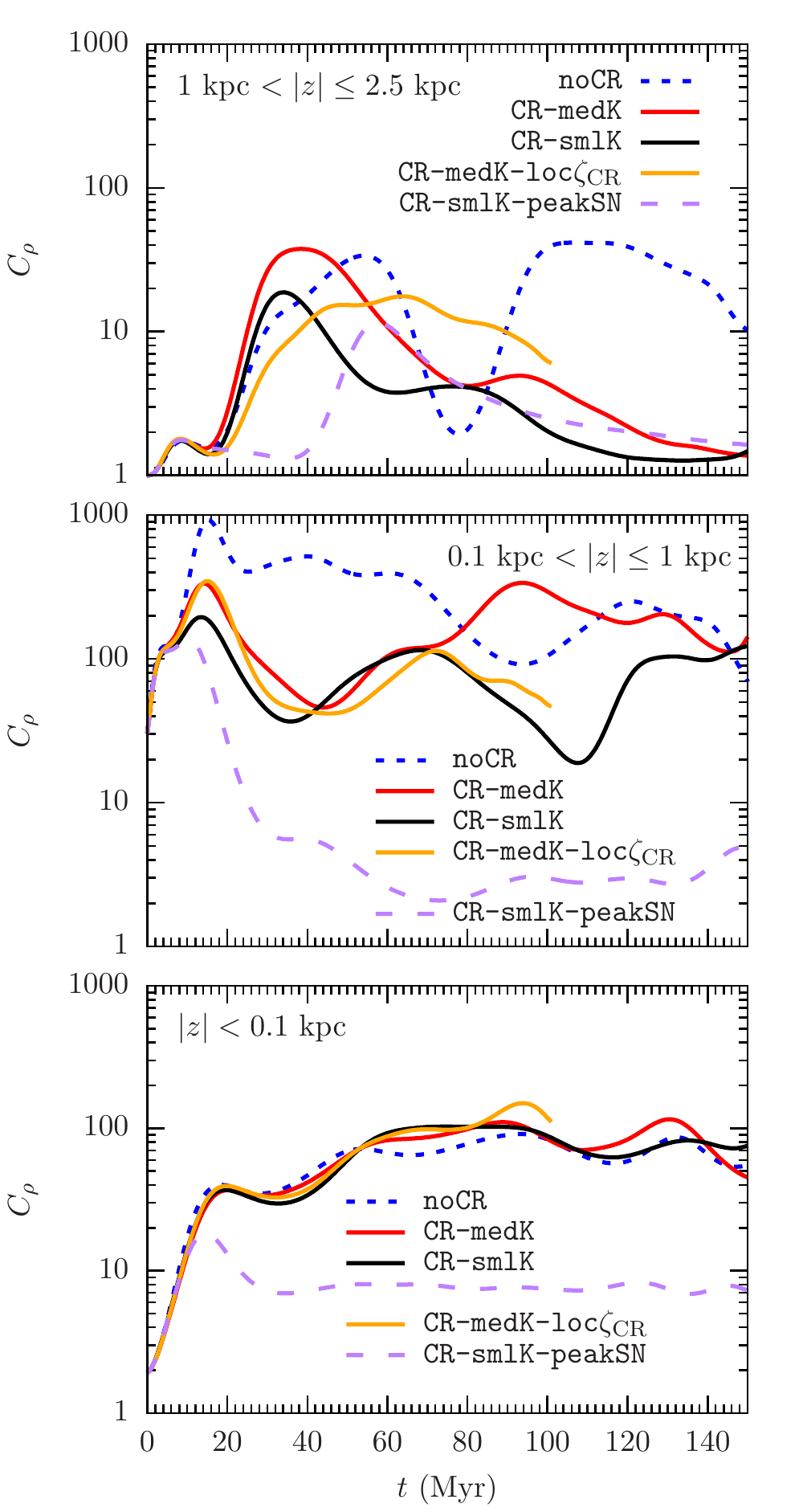}
   \caption{Clumping factor of the density for different heights. In the disc all simulations with clustered SN driving show a very similar evolution with clumping factors close to 100. The strongest clumping ($C_\rho\sim100$) occurs in the region from $0.1-1\,\mathrm{kpc}$ where the SNe explode in low-density environments and efficiently structure the outflowing gas. Above $z\sim1\,\mathrm{kpc}$ the gas becomes smoother ($C_\rho\sim10$) with very low values for the CR runs and strong temporal changes in the non-CR simulation.}
   \label{fig:clumping-factor-density}
\end{figure}

\begin{figure*}
\begin{minipage}{\textwidth}
  \includegraphics[width=\textwidth]{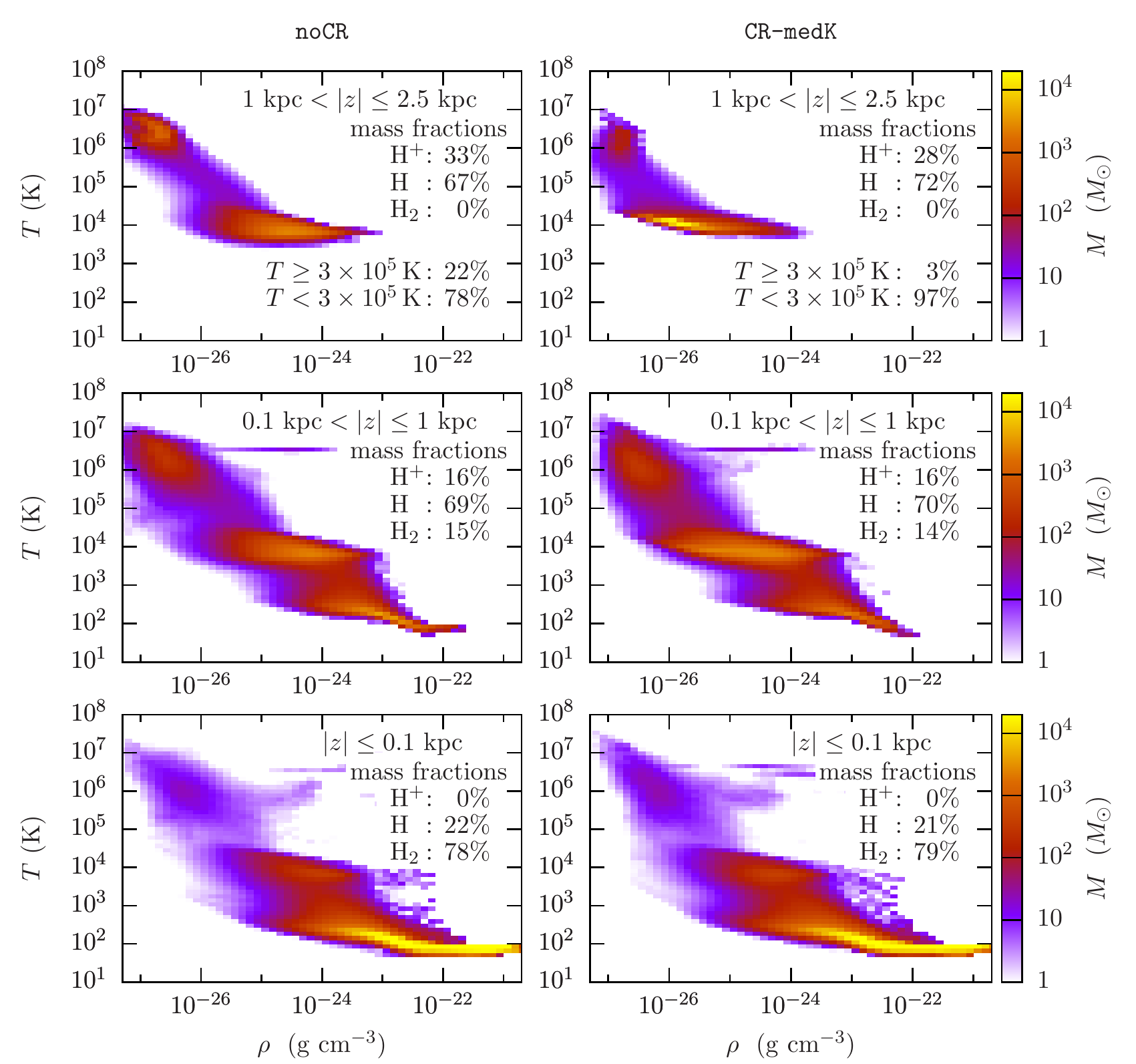}
\end{minipage}
  \caption{Two-dimensional histograms of the gas mass as a function of density and temperature for simulations \texttt{noCR} and \texttt{CR-medK} averaged from $t=50-100\,\mathrm{Myr}$ for different volumes of the simulation box. The time averaged distributions in the lower two volumes are similar for both runs. The upper volume that covers the outflow shows noticeable differences. In the thermal run 22\% of the gas is hot, whereas in the run including CRs only 3\% of the mass is above $3\times10^5\,\mathrm{K}$. Furthermore, the gas in run \texttt{CR-medK} is strongly concentrated around $\rho\sim10^{-26}\,\mathrm{g\,cm}^{-3}$ and $T\sim10^4\,\mathrm{K}$.}
  \label{fig:phase-plots-diff-heights}
\end{figure*}

\begin{figure}
  \includegraphics[width=8cm]{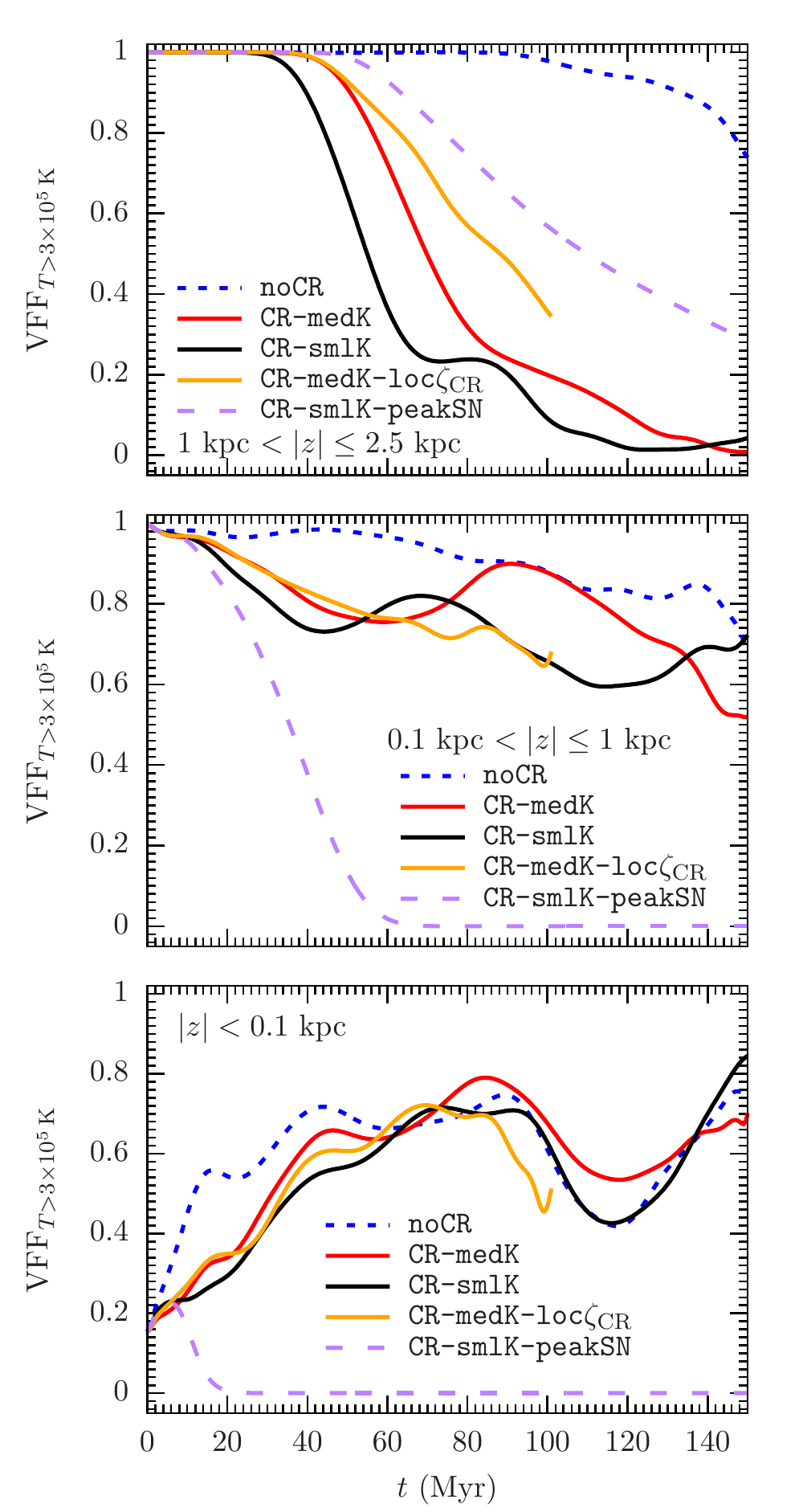}
  \caption{Volume filling fraction of the gas with a temperature of $T>3\times10^5\,\mathrm{K}$ over time at different heights. The disc as well as the lower halo are dominated by hot gas. At heights above $1\,\mathrm{kpc}$ the CR runs have very little gas above $T>3\times10^5\,\mathrm{K}$, in the purely thermal run most of the volume is hot. An exception is \texttt{CR-smlK-peakSN} where the disc and all outflowing gas is warm. CR driven outflows only fill the upper halo with warm gas after $\sim60-100\,\mathrm{Myr}$.}
  \label{fig:vff-hot-diff-heights}
\end{figure}

We expect a difference in the outflows depending on the CR diffusion coefficient. Intuitively, for a smaller diffusion coefficient it takes longer to reduce the CR pressure gradient. As a result the CRs have more time to accelerate the gas \emph{and} this acceleration is larger. Simulations by \citep{SalemBryan2014} confirm this theoretical considerations. On the other hand, an overdensity of CR energy also leads to higher CR cooling rates via hadronic cooling. However, as we discuss later this effect is subdominant.

The time evolution of vertical motions is shown in Fig.~\ref{fig:outflow-time-evol}. The left panels show the values at $1\,\mathrm{kpc}$ height, the right panels are at $2\,\mathrm{kpc}$. We compute mass flux rates at height $z$ as
\begin{equation}
\dot{M}_\mathrm{X} = \sum_i\rho_iv_{\mathrm{out},i}A_i,
\end{equation}
separately for inflowing ($\mathrm{X}=\mathrm{in}$) and outflowing gas ($\mathrm{X}=\mathrm{out}$) where $\rho_i$ is the cell density, $A_i$ the cell area $dx_i\times dy_i$, and $v_{\mathrm{out},i}=\mathrm{sgn}(z)\times v_{z,i}$ is the outward pointing velocity. The set $i$ includes cells that intersect with vertical positions $\pm z$. For outflow we only consider cells with $v_{\mathrm{out},i}>0$. Analogously, for inflow all cells with $v_{\mathrm{out},i}<0$ are taken into account. From top to bottom we show the mass flux rate, as well as the volume average of the density and the outflowing velocity at $\pm1\,\mathrm{kpc}$ (left) and $\pm2\,\mathrm{kpc}$ (right). All lines indicate the values for \emph{outflowing} gas, i.e. the values for cells $i$ with velocity vectors pointing away from the midplane. The dots connected with thin lines in the top panel are for the \emph{infalling} material. In the top panel we also indicate the star formation rate (SFR) which is directly connected to the SN rate (see introduction). The ratio of the outflow rate to the star formation rate is known as the mass-loading factor,
\begin{equation}
\eta=\frac{\dot{M}_\mathrm{out}}{\dot{M}_\mathrm{SFR}}.
\end{equation}
After $\sim40\,\mathrm{Myr}$ all simulations reach mass-loading factors of order unity at $1\,\mathrm{kpc}$ height. The mass flux shows fluctuations on short time scales. We therefore apply a $2\,\mathrm{Myr}$ averaging and smooth over small fluctuations. An exception is \texttt{CR-smlK-peakSN}, which needs longer to reach that height and shows very small temporal fluctuations around $\eta\approx1$ once the outflowing gas has reached the measurement height. Due to the temporal fluctuations it is difficult to rank the simulations according to their outflow efficiency and we will come back to it when discussing the gas masses in different volumes. After $\sim100\,\mathrm{Myr}$ the inflow at a height of $1\,\mathrm{kpc}$ starts to increase (dashed lines). For simulation \texttt{CR-smlK} the inflow is always below $\eta\sim0.1$, so there is net outflow over the entire simulation time. For \texttt{CR-medK} outflow and inflow approximately cancel after $t=110\,\mathrm{Myr}$, so the net gas flow stalls. This is consistent with the vanishing net velocities in Fig.~\ref{fig:vertical-profiles}. For the purely thermal simulation \texttt{noCR} the inflow starts to dominate in the last $15\,\mathrm{Myr}$, which is reflected in the negative velocity profile. At a height of $2\,\mathrm{kpc}$ the outflow rates also reach mass-loading factors of order unity but it takes $\sim50-70\,\mathrm{Myr}$. For \texttt{CR-smlK-peakSN} the gas barely reaches that height (cf. Fig.~\ref{fig:Hill-plots-10m-10sp-ng}). Towards the end of the simulation we note a clear difference between the thermal run that is again dominated by strong infall and the CR runs, in particular \texttt{CR-smlK}, which continues to drive a steady and strong outflow without any sign of relevant inflow or fountain activity.

The measured volume weighted densities and velocities are only plotted for the outflowing material in order to emphasise the difference in the driving process rather than the properties of infalling material. The curves indicate that in the CR runs the gas is a factor of a few denser ($\sim10^{-26}\,\mathrm{g~cm}^{-3}$) and a factor of a few slower ($30-40\,\mathrm{km~s}^{-1}$) compared to the thermally driven motions. This has also been found in our previous study \citep{GirichidisEtAl2016CR}, however the differences between non-CR runs and CRs are much smaller in this study. We attribute this to the four times higher spatial resolution. For the majority of the SNe the Sedov-Taylor radius is resolved with at least four cells. The reduced thermal over-cooling results in more efficient thermal SN driving in particular for the clustered SNe of type~II that explode in the disc region.

As the outflow rates vary over time and the net effect is difficult to extract, we show the time evolution of the gas masses in different volumes (see Fig.~\ref{fig:sketch-vertical-volumes}) in Fig.~\ref{fig:halo-mass-time-evol}. There is no infall from outside the simulation box, so all changes in mass with respect to the initial mass in the different sections of the box are due to outflows from the disc region or fountain flows falling back. The bottom panel shows the mass in the disc, which does not change so much relative to the initial mass and is therefore plotted in linear scale. The simulations clearly indicate the difference between CR supported SNe and purely thermal explosions. For times up to $100\,\mathrm{Myr}$ the CRs could remove roughly twice as much gas from the disc compared to the thermal run. The strongest outflows are measured for \texttt{CR-smlK-peakSN}, which is peculiar because the peak driving run without CRs does not drive any outflow (see Sec.~\ref{sec:discussion}). There is no significant difference between the CR runs with clustered SNe. In the last $40\,\mathrm{Myr}$ the thermal run can catch up pushing gas out of the disc region. However, the beginning infall of gas that has been lifted into the halo earlier suggests that this is only a temporary situation. The mass evolution in the disc is inversely visible in the volumes at larger heights, where we find overall higher masses if CRs are included.

\begin{table}
\caption{Average outflows and mass loading factors.}
\label{tab:effective-outflows}
\begin{tabular}{lccc}
simulation & $M_\mathrm{tot,>2.5\,\mathrm{kpc}}~(M_\odot)$ & $\langle\dot{M}\rangle_{100\,\mathrm{Myr}}~(M_\odot\,\mathrm{Myr}^{-1})$ & $\langle\eta\rangle$\\
\hline
\texttt{noCR} & $1.8\times10^4$ & $\phantom{0}180$ & $0.1$\\
\texttt{CR-medK} & $1.0\times10^5$ & $1022$ & $0.7$\\
\texttt{CR-smlK} & $2.1\times10^5$ & $2080$ & $1.4$\\
\hline
\end{tabular}

\medskip
Total masses of outflowing gas and effective outflows rates. Column~2 shows the total mass above $|z|=2.5\,\mathrm{kpc}$, column~3 the averaged outflow rate assuming that this mass has been accumulated during the last $100\,\mathrm{Myr}$ and column~4 the corresponding averaged mass loading factors.
\end{table}

Defining the effective outflow rates and mass loading factors are difficult because the simulations are not in steady state and the simulation time is not long enough to allow for an averaging over several fountain flow times. We can nonetheless take the halo mass above $|z|=2.5\,\mathrm{kpc}$, $M_\mathrm{tot,>2.5\,\mathrm{kpc}}$, at the final time of $t=150\,\mathrm{Myr}$ as a proxy for the outflows. Assuming that the gas needs a time of $\sim50\,\mathrm{Myr}$ to reach the height of $2.5\,\mathrm{kpc}$, we can compute an average outflow rate for the region above over the last $100\,\mathrm{Myr}$. We define $\langle\dot{M}\rangle_{100\,\mathrm{Myr}} = M_\mathrm{tot,>2.5\,\mathrm{kpc}}/100\,\mathrm{Myr}$ and the corresponding mass loading factor $\langle\eta\rangle = \langle\dot{M}\rangle_{100\,\mathrm{Myr}}/\mathrm{SFR}$. This averaging seems reasonable only for simulations \texttt{noCR}, \texttt{CR-medK}, and \texttt{CR-smlK} and the numbers are shown in table~\ref{tab:effective-outflows}. The total mass above $2.5\,\mathrm{kpc}$ varies strongly between the simulations with only $2\times10^4\,M_\odot$ for the purely thermal run and an order of magnitude more for \texttt{CR-smlK}, which is about 10\% of the total mass in the simulation box for the latter one. The averaged outflow rate over the last $100\,\mathrm{Myr}$ of evolution ranges from $\sim200-2000\,M_\odot\,\mathrm{Myr}^{-1}$. This corresponds to average mass loading factors of $0.1-1.4$. Although being very simplified estimates, the numbers support the paradigm that purely thermal SNe are only able to drive inefficient winds. If CRs are included the mass loading can increase to values of order unity.

\subsection{Clumping and phase structure}

The column density plots in Fig.~\ref{fig:Hill-plots-00-10s-ng} and \ref{fig:Hill-plots-10m-10sp-ng} visualise that the CR simulations generate a smooth halo region. We quantify the degree of density contrast in the gas using the clumping factor
\begin{equation}
	C_\rho = \frac{\langle\rho^2\rangle}{\langle\rho\rangle^2},
\end{equation}
which is commonly used in a cosmological context. Here, $\langle\cdot\rangle$ is the volume weighted average. Fig.~\ref{fig:clumping-factor-density} shows the time evolution of $C_\rho$ for different heights. In the disc the clumping is very similar for all simulations with clustered SNe with $C_\rho\approx50-100$, so the impact of CRs in shaping density contrasts around $z=0$ is very small. The region up to $1\,\mathrm{kpc}$ above the midplane is highly structured by the SNe with a higher clumping factor for the thermal run in the first half of the simulated time. Then the curves start overlapping with again values around 100. The variations over time for this section of the box are nicely illustrated by the time evolution in Fig.~\ref{fig:Hill-plots-00-10s-ng} and \ref{fig:Hill-plots-10m-10sp-ng}. In the range of $1-2.5\,\mathrm{kpc}$ the gas in the CR runs ends up at clumping factors of order unity, so the gas distribution does not just appear smooth in the column density plots. The few patches of gas that pass this volume in the purely thermal run result in strong temporal variations of $C_\rho$ ranging from a $1-40$, where the small values correspond to times, in which the halo basically consists of the low-density background gas rather than a smooth outflow. Simulation \texttt{CR-smlK-peakSN} marks again an exception with a smooth gas distributions everywhere. The broad peak of $C_\rho$ in the upper halo at $t\sim60\,\mathrm{Myr}$ marks the time when the outflowing gas enters this region of the simulation box.

The clumping of the gas is connected to the temperatures in the box. Fig.~\ref{fig:phase-plots-diff-heights} depicts density-temperature phase plots for simulations \texttt{noCR} (left) and \texttt{CR-medK} (right) at different heights (top to bottom). Colour-coded is the mass. In addition we show the relative hydrogen mass fractions in each analysis volume as well as the mass fractions of the hot gas with $T>3\times10^5\,\mathrm{K}$ for the upper halo, which contains most of the outflowing gas. The plots and numbers are averaged quantities from $t=50-150\,\mathrm{Myr}$. In the disc and the lower halo the averaged phase plots are very similar for both runs. In the upper volume we note three important differences in the case for \texttt{CR-medK}. The first is the approximately 60\% larger total mass. The second is the stronger concentration of the mass at densities of $\sim10^{-26}\,\mathrm{g\,cm}^{-3}$ and temperatures of $\sim10^4\,\mathrm{K}$. And finally the fraction of hot mass in the upper halo. Whereas in \texttt{noCR} 22\% of mass is hot with $T>3\times10^5\,\mathrm{K}$, the fraction in run \texttt{CR-medK} is only 3\%. For simulation \texttt{CR-smlK} (not shown) the distributions and mass fractions are very similar. The chemical composition does not differ perceptibly between the two runs.

We complement the phase plots with the time evolution of the volume filling factor of the hot gas, $\mathrm{VFF}$, for temperatures above $3\times10^5\,\mathrm{K}$ in Fig.~\ref{fig:vff-hot-diff-heights}. From top to bottom we subdivide the volume into the usual regions. In the disc the SNe create hot density voids, which occupy $60-80\%$ of the volume. The temporal variations are much stronger than the differences between non-CR and CR runs. The activation of a few new SN clusters in dense gas are visible in the decrease at $t\sim100\,\mathrm{Myr}$. After the first few SNe in these clusters have cleared the dense region the fraction of hot gas increases again. The volumes, in which SNe directly structure the medium ($0.1-1\,\mathrm{kpc}$), are dominated by hot gas with VFFs of $0.6-0.8$. The temporal variations are again large, however there is a tendency of the purely thermal run to have larger VFF than the runs including CRs. In the upper volume ($1-2.5\,\mathrm{kpc}$) we notice the strongest difference between the CR supported outflowing gas and the gas that is lifted by thermal SNe. Whereas simulation \texttt{noCR} has a VFF of the hot gas larger than $80-90\%$, the halo gas in the CR runs is warm with the majority of the volume occupied with gas colder than $3\times10^5\,\mathrm{K}$. For the simulation with small diffusion coefficient the density of the outflow is largest and the smooth gas results in the fastest population of the halo with colder gas. Again the simulation with peak driving is an exception with no hot gas in the disc and the lower halo as soon as the warm outflows reach the corresponding heights.

\subsection{Relative energy densities}

\begin{figure}
  \includegraphics[width=8cm]{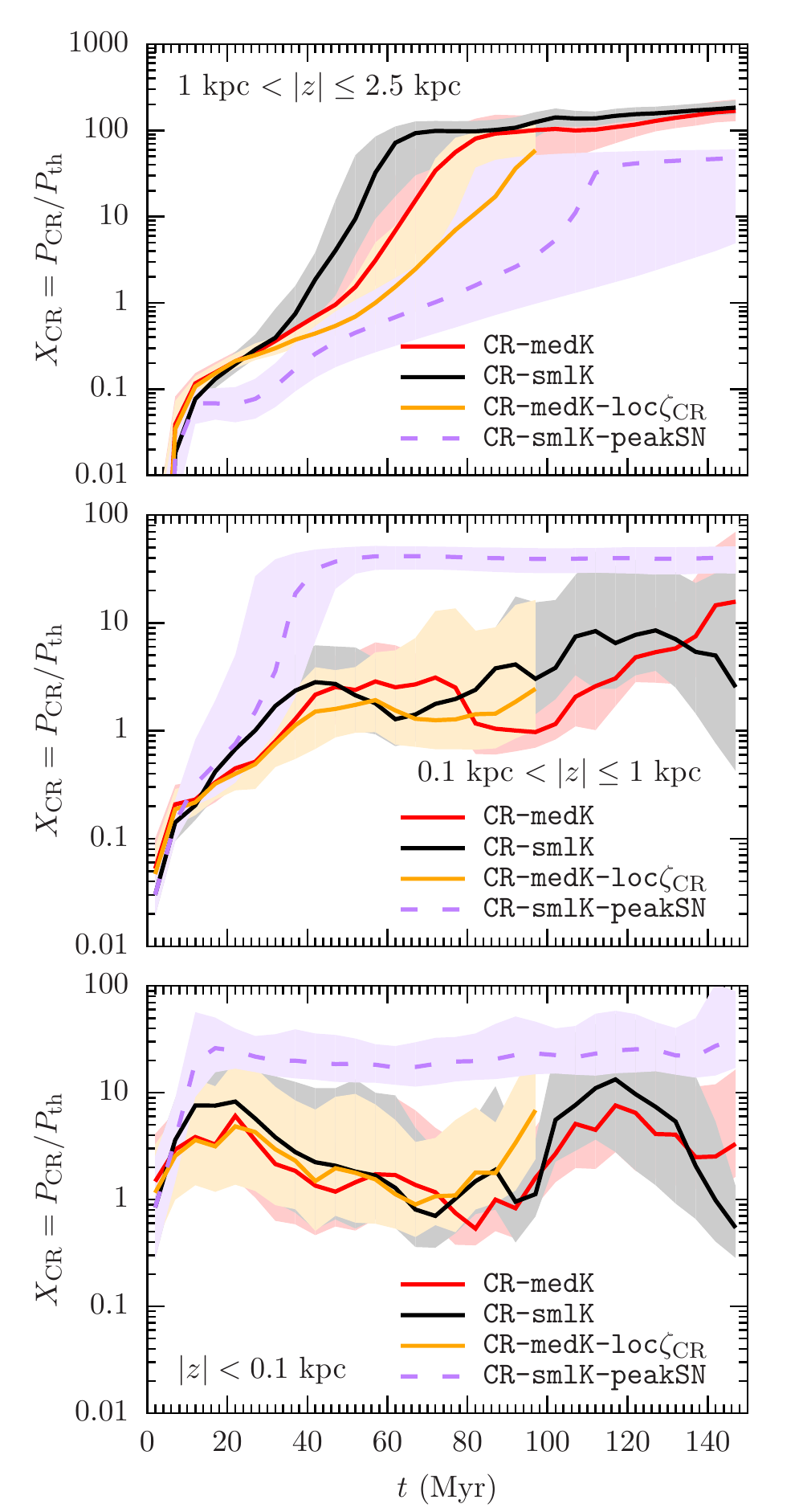}
  \caption{Time evolution of $X_\mathrm{CR}=p_\mathrm{CR}/p_\mathrm{th}$ in different volumes. The solid lines show the volume weighted median (see text) together with the 25 and 75 percentile of the distribution (shaded area). In the disc the CR pressure slightly dominates. The hot gas in the middle volume allows for regions of pressure equilibrium. At large heights CR dominate by far.}
  \label{fig:energy-ratios-time-evol}
\end{figure}

Fig.~\ref{fig:energy-ratios-time-evol} shows the time evolution of the CR to thermal energy ratio $X_\mathrm{CR}=P_\mathrm{CR}/P_\mathrm{th}$. From top to bottom we distinguish again between the different volumes previously defined. The solid lines show the volume weighted median of the distribution ($q=0.5$), which is the element $X_j$ of the sorted list of cell based ($i$) ratios $X_i$ that satisfies
\begin{equation}
\sum_{i=1}^{j}\,\frac{V_i}{V_\mathrm{tot}}\le q,\quad \text{and}\quad \sum_{i=j+1}^{N}\,\frac{V_i}{V_\mathrm{tot}}\le 1-q.
\end{equation}
Here $V_i$ are the corresponding cell volumes for ratio $X_i$ and $V_\mathrm{tot}$ is the total volume of the region of interest. The shaded area is bounded by the 25 and 75 percentile of the distribution ($q=0.25$ and $q=0.75$).

In the disc the values range from $\sim1-10$ over time with negligible difference between the individual runs. The width of the distribution covers around one order of magnitude. The volume above the disc is initially dominated by thermal pressure. In the time window from $\sim40-100\,\mathrm{Myr}$ we note an approximate pressure equilibrium. Towards the end of the simulation the ratio increases further to values of $X_\mathrm{CR}\sim10$, again with larger temporal variations than the ones between the simulations. In the upper halo the CR pressure clearly dominates over the thermal energy by several orders of magnitude approaching $X_\mathrm{CR}\sim200$ at the end of the simulated time. This is not surprising because the halo is filled by warm rather than hot gas, so the thermal pressure is comparably low. The density however is not high enough to cool away the CRs over the simulated time. The mass weighted distributions (not shown) are similar in the disc and the region above $|z|=1\,\mathrm{kpc}$. In the lower halo the values are larger by a factor of a few. The low temperature in run \texttt{CR-smlK-peakSN} results in large values of $X_\mathrm{CR}\sim20-70$ in all parts of the simulation box. The small variations over a long period of time indicate that there is an equilibrium between the CR injection and the CR losses.

\section{CR energy distribution, hadronic cooling and Gamma-ray emission}

\begin{figure*}
\begin{minipage}{\textwidth}
\centering
\includegraphics[width=14cm]{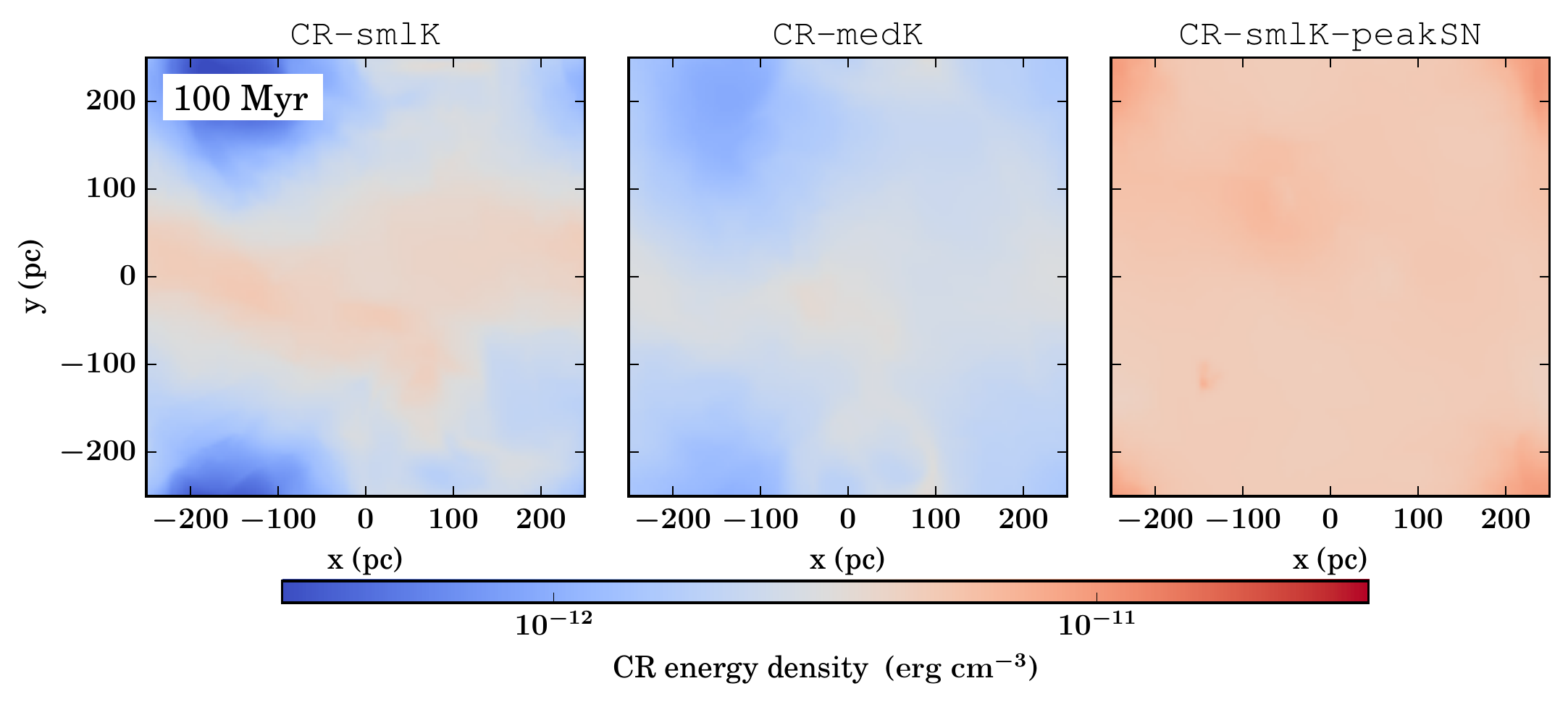}
\end{minipage}
\caption{CR energy density in the midplane for simulations \texttt{CR-smlK} (left), \texttt{CR-medK} (centre) and \texttt{CR-smlK-peakSN} (right). The variations are stronger for the smaller diffusion coefficient, but in both cases less than an order of magnitude. The spatial scales over which the energy changes are of the order of the box size. The spatial variations in simulation \texttt{CR-smlK-peakSN} are also small with overall larger CR energy densities.}
\label{fig:encrvol-midplane}
\end{figure*}

\begin{figure*}
\begin{minipage}{\textwidth}
\includegraphics[width=\textwidth]{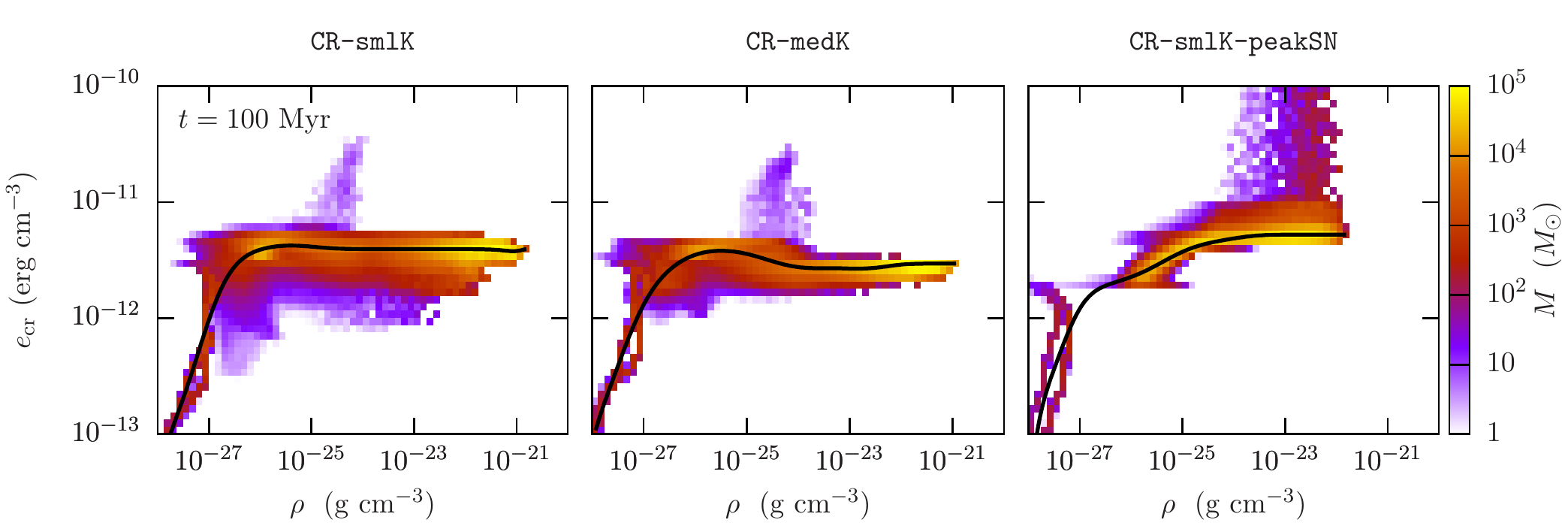}
\end{minipage}
\caption{Two-dimensional histograms of the mass as a function of gas density and CR energy density for simulations \texttt{CR-smlK} (legft), \texttt{CR-medK} (middle) and \texttt{CR-smlK-peakSN} (right) at $100\,\mathrm{Myr}$. The median of the distribution is shown as solid line. The distributions are very narrow in the CR energy density. \texttt{CR-smlK-peakSN} shows a weak positive trend of $e_{_\mathrm{CR}}$ with $\rho$. The low-energy wing is at the lowest densities and corresponds to the (almost) pristine gas at several kpc height. The patches of high $e_{_\mathrm{CR}}$ correspond to recent SN events, which are launched at intermediate densities in \texttt{CR-smlK} and \texttt{CR-medK} and at high densities in \texttt{CR-smlK-peakSN}.}
\label{fig:phaseplot-dens-encr}
\end{figure*}

\begin{figure}
\includegraphics[width=8cm]{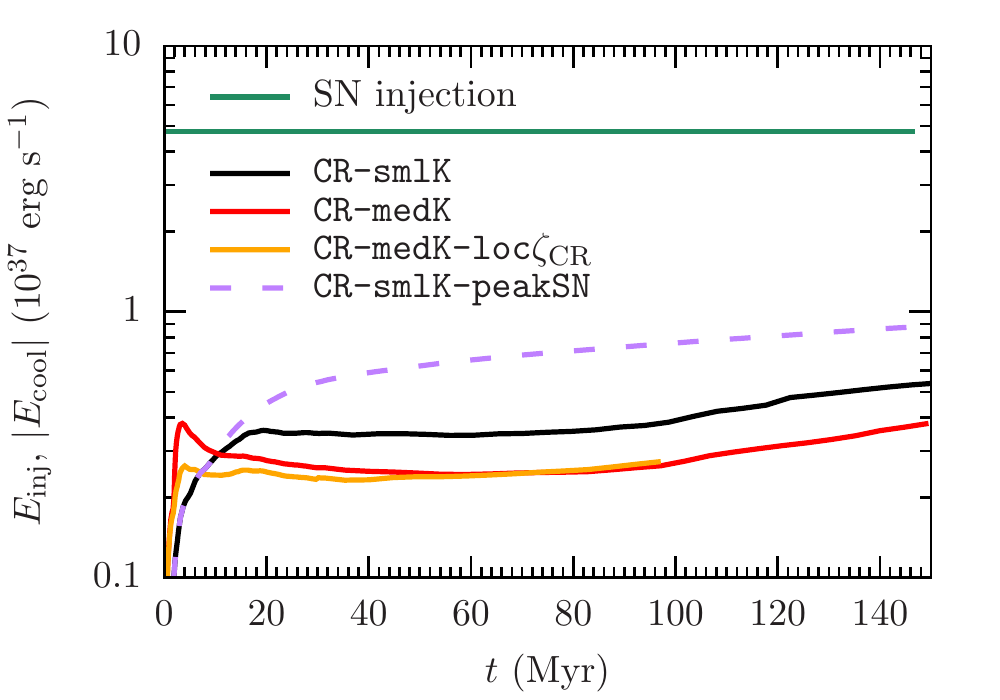}
\caption{CR injection rate (green line) and time averaged hadronic losses (black, red, blue, and purple lines). A fraction of $5-25\%$ of all the CR energy is lost via hadronic interactions. Lower diffusion coefficients (\texttt{CR-smlK}) as well as CR injection in the densest regions (\texttt{CR-smlK-peakSN}) result in longer time scales for the CRs to escape from the dense regions of the disc. This leads to a higher loss rates.}
\label{fig:CRinjection-hadronic-cooling}
\end{figure}

\begin{figure}
\includegraphics[width=8cm]{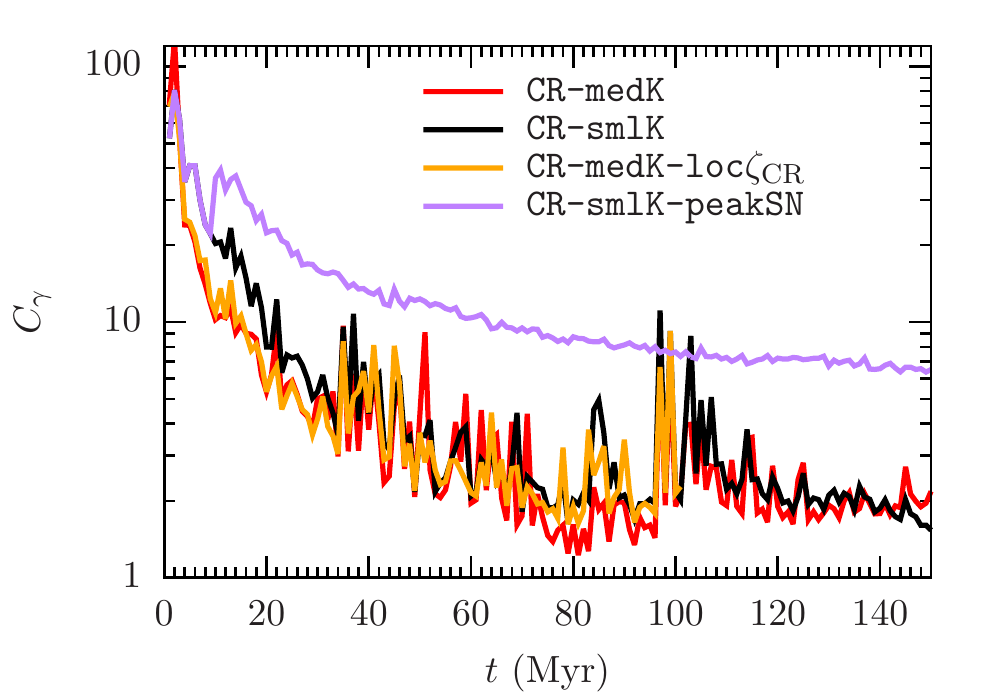}
\caption{Gamma-ray clumping factor for the total simulation box. In the beginning the CR distribution and the gas differ perceptibly, which results in a large clumping factor. Over time, the combined distribution becomes more uniform. The converged values between $2$ and $6$ indicate an overall positive correlation of gas and CRs.}
\label{fig:gamma-ray-clumping-factor}
\end{figure}

The hadronic losses scale with the gas density and the CR energy density (Eq.~\ref{eq:hadr-losses}), so the correlation of both quantities is of interest for CR-driven dynamics. As most of the hadronic losses are emitted via $\gamma$-ray photons, both the total losses as well as the locations of the losses are instructive measures. Whereas the density varies by up to eight orders of magnitude, the CR energy density is rather smooth. This is because CRs quickly diffuse through the ISM efficiently reducing local overdensities in their energy distribution. We show the energy density in the midplane in Fig.~\ref{fig:encrvol-midplane} for simulations \texttt{CR-smlK} (left), \texttt{CR-medK} (centre) and \texttt{CR-smlK-peakSN} (right) at $t=100\,\mathrm{Myr}$. In all simulations the range of the CR energy density is about an order of magnitude with the largest variations for the simulation with small CR diffusion coefficient. The spatial scales over which the CR energy density changes from a CR overdensity to an underdensity is of the order of the box length, i.e. the two-dimensional power spectra of the $e_{_\mathrm{CR}}$ field peak at the largest mode (not shown).

The correlation with the density is shown in the two-dimensional histograms in Fig.~\ref{fig:phaseplot-dens-encr} for simulations \texttt{CR-smlK}, \texttt{CR-medK} and \texttt{CR-smlK-peakSN} at $t=100\,\mathrm{Myr}$. Colour coded is the mass. The black line denotes the median of the distribution along the ordinate. Here we take all the gas in the simulation box into account. The narrow distributions are very similar for all runs with the majority of the mass located at $e_{_\mathrm{CR}}\sim3-5\times10^{-12}\,\mathrm{erg\,cm}^{-3}$. The correlation between CR energy density and gas density is very flat. Only simulation \texttt{CR-smlK-peakSN} shows higher CR energies at higher densities, which is expected as all SNe explode in the densest regions of the box. The extent in CR energy above $10^{-11}\,\mathrm{erg\,cm}^{-3}$ is caused by SN injections and quickly vanishes due to fast diffusion. For simulations \texttt{CR-smlK} and \texttt{CR-medK} many SNe explode in low-density regions and create CR overdensities in density voids. Therefore, we note a weak temporal anti-correlation which is visible in the middle panel. For simulation \texttt{CR-smlK-peakSN} the CR injection in dense gas leads to a weak positive correlation of $e_{_\mathrm{CR}}$ with $\rho$.

Fig.~\ref{fig:CRinjection-hadronic-cooling} indicates how much of the CR energy is lost due to hadronic cooling. Plotted are the CR injection rate by SNe as well as the cooling losses for the individual simulations. The injection is constant for all simulations because of the constant SN rate and the same CR efficiency per SN. The hadronic cooling rate varies from $5-25\%$ of the injection rate, where the lowest loss rates are for the simulations with a higher diffusion coefficient (\texttt{CR-medK}, \texttt{CR-medK-loc}$\zeta_\mathrm{CR}$). The effects of a locally varying CR ionisation rate are negligible. For the simulation with a smaller diffusion coefficient (\texttt{CR-smlK}), the CRs need more time to escape the disc region which allows them to cool more than in the simulations with larger diffusion coefficient. If the CRs are injected in dense regions their cooling rate is expected to be higher, which is confirmed by the numbers for run \texttt{CR-smlK-peakSN}.

Analogously to the density clumping, we analyse the correlation between the gas density and the CR energy density using a (hadronic / $\gamma$-ray) clumping factor
\begin{equation}
C_\gamma = \frac{\langle e_{_\mathrm{CR}}\rho\rangle}{\langle e_{_\mathrm{CR}}\rangle\langle\rho\rangle},
\end{equation}
where $\langle\cdot\rangle$ is the volume average over the entire domain. The time evolution of $C_\gamma$ is shown in Fig.~\ref{fig:gamma-ray-clumping-factor}. Initially, CRs and gas are correlated, which manifestst in large clumping factors. As CRs diffuse and outflows develop the clumping decreases. After $\sim100\,\mathrm{Myr}$ an overall steady distribution establishes, which indicates an equilibrium in the CR-gas correlation. For \texttt{CR-smlK-peakSN} the clumping is higher with $C_\gamma\approx6-8$ and small variations over time. For runs \texttt{CR-smlK}, \texttt{CR-medK} and \texttt{CR-medK-loc}$\zeta_\mathrm{CR}$ we find smaller values of $C_\gamma\approx2-3$ with temporally large peaks up to $C_\gamma\sim10$, which are attributed to SNe in dense environments. We note that the correlation over the entire box is always positive ($C_\gamma>1$), which reflects the vertical stratification of both $\rho$ and $e_{_\mathrm{CR}}$.

\section{Discussion and Caveats}
\label{sec:discussion} 
 
\subsection{Diffuse gas in the disc and the impact of CRs}

The outflows generated in our simulations are a combination of thermal and CR feedback by SNe. Locally at the SN positions the thermal feedback dynamically dominates by far, i.e. the thermal pressure gradients dominate,  because the injected fraction of CRs is only 10\% of the thermal energy. In dense regions the thermal energy can be quickly radiated away and the thermal pressure drops. In simulations where SNe are placed in density peaks the strong cooling can severely affect the generation of outflows. In our previous models the peak driving mechanism lead to a smooth density distribution in the disc with only warm diffuse gas in the midplane \citep{SILCC1}. The formation of dense clouds was prevented by SNe that are placed in the densest regions. The strong cooling in turn lead to inefficient heating by SNe, which prevented the formation of hot gas in the disc resulting in negligible volume filling fractions for gas with $T>3\times10^5\,\mathrm{K}$. As volume filling hot gas is crucial for thermally driven outflows \citep{NaabOstriker2017}, there was no thermal pressure support to launch an outflows from the disc in \citet{GirichidisSILCC2}. Including CRs, the thermal pressure is lost equally fast (see temperature profiles in Fig.~\ref{fig:vertical-profiles} and volume filling fractions in Fig.~\ref{fig:vff-hot-diff-heights}). As the CRs are injected together with the thermal energy in dense regions the CR losses are also higher in the run with peak driving. However, the total CR loss rates are only up to $25\%$ of the injected energy and the remaining $75\%$ are available for building up a pressure gradient and driving an outflow. The fact that the gas is smooth actually provides the largest effective cross section for the CRs. As the CR pressure gradient is very shallow, the non-thermal fluid cannot efficiently accelerate gas that is bound to dense clouds and thus locked. It is therefore not surprising that the densest outflows are driven in simulation \texttt{CR-smlK-peakSN}. As the thermal support for driving the gas out of the disc is negligible, CRs alone are able to drive and sustain a smooth outflow with a mass loading factor of order unity. This is in good agreement with the results in \citet{GirichidisEtAl2016CR} with four times coarser resolution.

Our results agree qualitatively with the simulations by \citet{SimpsonEtAl2016} in a similar setup with using CR solver in \textsc{Arepo} \citep{Springel2010, PakmorEtAl2016, PfrommerEtAl2017}. They use a Schmidt-type star formation rate \citep{Schmidt1959}, which is converted to an instantaneous SN rate. This approach is very similar to our peak driving model except for the fact that their SN rate varies based on the density structure in the disc.

\subsection{Diffuse ionised gas layer above the disc}

Observations of the Milky Way and other galaxies show an extended layer of diffuse ionised gas above the disc with scale heights of more than a kiloparsec \citep{HaffnerEtAl2009}. This warm ($T\sim10^4\,\mathrm{K}$) gas is most likely ionised by O and B stars in the disc \citep{Reynolds1990, Reynolds1990b}. Numerical modelling of photoionisation support the scenario that O stars can be responsible for this layer of ionised gas if the ISM is porous and provides low-density channels through which the photons can travel \citep{WoodEtAl2005, WoodEtAl2010, BarnesEtAl2014}. Both the fact that gas is warm and can be ionised by O and B stars suggests that it might not be driven out of the disc by hot ($T>10^6\,\mathrm{K}$) SN remnants. Instead a more natural explanation would be a non-thermal support. Previous numerical models of feedback driven outflows have difficulties to create this layer with the observed scale heights. \citet{HillEtAl2012} performed numerical simulations of the magnetised ISM in a setup similar to the one in our study. They ruled out magnetic fields as the responsible mechanism to support a vertically extended gas distribution with the observed scale height. The CR driven gas in this study is predominantly warm and slowly lifted into the halo, which qualitatively results in an extended distribution of warm gas and might thus serve as a potential explanation of that layer.

We do not include ionising sources in the present work, neither CR heating of the warm ionised medium \citep{WienerZweibelOh2013}, which results in mostly warm neutral gas in the extended layer around the disc. Both processes have been accounted for in \citet{VandenbrouckeEtAl2018}, who perform a post-processing analysis of the simulations by \citet{GirichidisEtAl2016CR} with Monte-Carlo radiative transfer. %Using the Monte Carlo radiative transfer code \textsc{CMacIonize} \citep{VandenbrouckeWood2017}
Their computed vertical profiles of the diffuse ionised gas are in good agreement with the observations, which emphasises the role of CRs in forming the extended layer of warm ionised gas.

\citet{PetersEtAl2015} post-process the simulations by \citet{GirichidisEtAl2016CR} with a focus on the surface brightness in soft X-ray. The emission coming from the warm neutral gas layer is about one order of magnitude below the median of the observed values of the Milky Way halo by \citet{HenleyShelton2013}. The better resolution and the more efficient SN heating in the present study are likely to correct the surface brightness in the right direction.

\subsection{Missing CR physics}

We would like to discuss two important aspects of missing CR physics, namely the missing loss processes and the diffusion limit we apply. Concerning the losses we need to distinguish between different energy ranges for the CRs. For low-energy CRs with momenta of $p_{_\mathrm{CR}}\lesssim0.1\,\mathrm{GeV}/c$ Coulomb losses dominate the energetic evolution. However, without following the energy spectrally resolved the loss rates due to Coulomb interactions will result in overall too efficient losses because not only the low-energy part of the spectrum will be cooled away. Instead, the lost energy due to Coulomb cooling using an assumed fixed spectral shape will lead to cooling in effectively all energy ranges.

We use the advection-diffusion approximation for the CR transport. However, the CR transport process depends on several environmental factors in which the commonly used approximation is a strong simplification. The review by \citet{Zweibel2013} outlines the two most important treatments, namely \emph{self-confinement} and \emph{extrinsic turbulence}. In the self-confinement picture CRs efficiently amplify Alfv\'{e}n waves if they are in resonance with the gyration (gyroresonant scattering, \citealt{KulsrudPearce1969}, \citealt{Wentzel1974}). If the excited waves are not efficiently damped the CRs stream with the Alfv\'{e}n velocity. Damping processes like turbulent damping, ion-neutral damping or nonlinear Landau damping depend on the thermal and turbulent state of the gas and regulate the streaming velocity \citep{KulsrudCesarsky1971, FarmerGoldreich2004, WienerOhGuo2013}.

In case of extrinsic turbulence the turbulent cascade or other energy injections are the main source of waves that the CRs can scatter off. Those waves isotropise the CR distribution and couple them to the gas. If the waves are predominently Alfv\'{e}n waves the travelling speed scales again with the Alfv\'{e}n velocity modulated by the scattering frequency.

Most likely the simple diffusion approximation is valid for high energy CRs ($\gtrsim100\,\mathrm{GeV}$) whose total energy density is negligible compared to the other energies (magnetic, thermal and kinetic). For CRs at the bulge of the total CR energy with particle momenta of a few GeV/$c$ the resonant scattering might be non-negligible and streaming effects need to be taken into account. \citet{UhligEtAl2012} included CR streaming into simulations of Galaxy formation. More recently, \citet{RuszkowskiYangZweibel2017} simulated isolated galaxies including anisotropic CR diffusion and streaming. Although a detailed comparison of the CR effects related to streaming compared to the simple diffusion approximation is difficult for us, the overall impact of CRs found by the latter authors are well in line with our results.

In the diffusion approximation the diffusion coefficient depends on the CR energy. The gyroradius of CRs depends linearly on the particle momentum, which results in larger mean free paths and consequently in larger diffusion coefficients, which scale as $\mathsf{K}\propto E^s$, where $s$ ranges from $0.3-0.6$ \citep{StrongMoskalenkoPtuskin2007}. The fiducial value of $10^{28}\,\mathrm{cm}^2\,\mathrm{s}^{-1}$ applies to CRs in the GeV range. The limitations of using a single energy description of CRs and one fixed energy for the diffusion coefficient vary for different energy ranges. As the CR spectrum peaks at an energy of a few GeV and steeply declines for higher and lower particle energies, the bulk of the CR energy is deposited in a relatively narrow energy regime around a few GeV. The corresponding variations of the diffusion coefficient are of the order of a few, which is likely to change details of the outflows but is unlikely to fundamentally alter the impact of CRs on the ISM (see differences between \texttt{CR-smlK} and \texttt{CR-medK}). Low energy CRs with energies of a few MeV diffuse at slower speeds by $1-2$ orders of magnitude. The resulting CR ionisation rates in the dense regions of the ISM are likely to be significantly higher compared to our approach in \texttt{CR-medK-loc}$\zeta_\mathrm{CR}$. The total amount of molecular hydrogen might be affected but we do not expect the global dynamics to be significantly altered. For energies above the spectral peak the faster diffusion will result in fewer high-energy CRs in the disc region. The hadronic losses are therefore likely to decrease. However, given the small loss rates and the steep spectrum for high-energy CRs we again do not expect the dynamics our simulations to differ perceptibly.

\subsection{Magnetic field properties}

We overall find weaker magnetic field in our simulations compared to observed values. Our setup is missing the effects of shear due to differential rotation of the disc and the resulting amplification as well as the alignment of the field preferentially parallel to the disc. This field geometry would reduce the effective diffusion of CRs into the halo. One the one hand there would be stronger CR pressure gradients that might enhance the outflows. On the other hand the CRs would remain longer in dense environments and the effective losses due to hadronic interactions would be larger.

\citet{PakmorEtAl2016} has shown that a fast diffusion of CRs into the halo can efficiently reduce the magnetic field amplification. The weak fields in our simulations might thus not only be due to the relatively low resolution and the resulting weak turbulent dynamo. Which of the effects will be dominant cannot simply be answered with our current setup.

A more detailed study on CR-driven galactic dynamos has been performed by \citet{HanaszEtAl2009} in a similar setup of a stratified box. The study demonstrates that the  horizontal magnetic field resulting from the CR-driven dynamo dominates for a wide range of model parameters. The field geometry is controlled by the value of magnetic diffusivity, the value of parallel and perpendicular diffusion coefficients as well as the supernova rate. However, even in cases of efficient amplification of a horizontal field component up to equipartition with the kinetic energy, the vertical, randomly oriented field may dominate if the dissipation of vertical field by magnetic diffusivity is slower than the generation by SNe. In all of their simulations the CR energy dominates over kinetic and magnetic energies by $1-1.5$ orders of magnitude, even if the magnetic field reaches equipartition with the kinetic energy, contrary to what is deduced from observations. The relatively large CR energy density might be related to the geometry of stratified box models. With periodic boundary conditions in $x$ and $y$ the CR energy can only decrease by loss processes or diffusion through the $z$ boundary. In a global galactic setup the spiral structure of a full disc allows for a more versatile distribution of the CR energy.

With global magnetic field energy densities that are significantly below the CR counterpart, the dynamical coupling between CRs and magnetic field is expected to be stronger than the simplified diffusion approach applied in the current simulations. The weak fields would not be able to contain the CRs in the midplane. To properly account for the dynamical interplay, the magnetic field would need to be coupled to the CRs with a more self-consistently treatment of the relevant plasma processes.

\subsection{Simulating outflows}

ISM models and outflows have been studied in similar setups with different foci and complexity \citep{CreaseyTheunsBower2013, CreaseyTheunsBower2015, GirichidisSILCC2, MartizziEtAl2016, SimpsonEtAl2016, LiBryanOstriker2017, GattoEtAl2017}. Most of these studies include SNe as feedback from stars. The simulations by \citet{CreaseyTheunsBower2013} barely reach mass loading factors of order unity for a large range of gas surface densities using SN feedback at a constant rate. \citet{MartizziEtAl2016} change disc model and SN rate with generally small mass loading of less than unity for most of their runs, measured at e.g. $500\,\mathrm{pc}$. \citet{LiBryanOstriker2017} use a similar SN driving schemes as in this work with different distributions for type~Ia and type~II SNe. Their outflows exceed loading factors of unity for their MW conditions even at a height of $1\,\mathrm{kpc}$. The volume filling fraction of the hot gas is 20\% in their MW simulation, which is in good agreement with our lower halo properties. At larger heights, only simulation \texttt{noCR} can maintain a similarly large VFF of the hot gas. All simulations with CRs evolve towards significantly lower fractions. 

The consistent formation of star clusters using sink particles and a correlated feedback are tested in \citet{GattoEtAl2017} and \citet{KimOstriker2018}. Both studies include the dynamical formation of sink particles as star clusters, accretion of gas and (delayed) feedback based on stellar evolution models. \citet{GattoEtAl2017} also include winds from massive stars, \citet{KimOstriker2018} account for a runaway O-star component based on the sink particle properties. \citet{GattoEtAl2017} find that both the SNe alone as well as the combination of SNe and winds are able to launch strong outflows with mass loading factors above unity once the volume filling fraction of the hot gas in the disc ($|z|<100\,\mathrm{pc}$) exceeds 50\%. \citet{KimOstriker2018} simulate the evolution of the ISM for about $600\,\mathrm{Myr}$ including galactic shear, which allows them to investigate the long-term behaviour of the gas cycle and fountain flows. They find periodic cycles which repeat after approximately $50\,\mathrm{Myr}$. Outflow-dominated phases show large volume-filling fractions of hot gas ($\sim10^6\,\mathrm{K}$). During inflow-dominated times the gas in the halo is warm with temperatures between $10^4$ and $10^5\,\mathrm{K}$. 

\citet{SimpsonEtAl2016} follow a locally and temporally varying star formation rate based on \citet{Schmidt1959}, which is converted to an instantaneous SN rate. They also include CRs in a very similar manner as in our study and follow the thermal and chemical evolution of the ISM with the same chemical network. Their model without CRs does not drive any measurable outflow due to the efficient cooling of the SN remnants. Including CRs the mass loading factors reach unity, where the outflow rate is computed as the loss rate through the boundary of the box at $\pm5\,\mathrm{kpc}$. The initial delay until the halo gas reaches that height is likely to result in higher outflow rates if measured at lower altitudes, in particular as the star formation rate tends to slightly decrease throughout the simulation time whereas the outflow rate is slowly increasing.

\citet{FarberEtAl2018} investigate CR supported outflows in a very similar setup as our box including a temperature dependent diffusion coefficient to mimic the CR coupling to the gas depending on the effective ionisation state. At temperatures above $10^4\,\mathrm{K}$ they reduce the parallel diffusion coefficient to $3\times10^{27}\,\mathrm{cm}^2\,\mathrm{s}^{-1}$, one order of magnitude lower than the fiducial Galactic value. In regions with $T<10^4\,\mathrm{K}$ they switch to a larger parallel diffusion coefficient of $10^{29}\,\mathrm{cm}^2\,\mathrm{s}^{-1}$. The CR supported outflows in their simulations show very similar temperature structures as in our run with dense outflows ($\rho\sim3\times10^{-27}-2\times10^{-25}\,\mathrm{g\,cm}^{-3}$) and temperatures below $10^5\,\mathrm{K}$.

The impact of CRs has also been investigated on larger scales, i.e. on full galactic discs. Given the large dynamic range, it is infeasible to model large-scales dynamics and ISM details at the complexity described in the studies above at the same time. \citet{HanaszEtAl2013} use an isothermal equation of state and demonstrate that CRs alone are able to launch winds from the disc to altitudes of $40\,\mathrm{kpc}$ with mass loading factors of order unity even at $10\,\mathrm{kpc}$ height. \citet{BoothEtAl2013} investigate CRs in MW and SMC analogues finding  no relevant difference between purely thermal and CR supported winds	 in MW-like conditions. In their SMC model CRs are able to increase the mass loading factor by a few to $\eta\sim1-10$ measured at $z=20\,\mathrm{kpc}$. \citet{SalemBryan2014} perform a large set of simulations with varying diffusion coefficient, CR injection fraction and finding $\eta=0.3$ for their fiducial MW-like run, decreasing with increasing $\mathsf{K}$. More recently, \citet{PakmorEtAl2016} compare isotropic with anisotropic diffusion and report that the latter mode is in much better agreement with observed discs. 

Our CR assisted outflows are slow on scales of $\sim1-2\,\mathrm{kpc}$ above the disc, which is much less than the escape velocity for local environments, e.g. $v_\mathrm{esc}\sim300$ for the MW at the solar radius. However, the winds are further accelerated at larger heights and thus do not need to be accelerated above escape velocity at low altitudes. If the CRs provide an additional smooth acceleration that is opposing gravity, the gas can be slowly lifted rather than being sped up above escape velocity. Detailed one-dimensional models by \citet{BreitschwerdtMcKenzieVoelk1991} and \citet{DorfiBreitschwerdt2012} find steady wind solutions, in which the gas is continuously accelerated even at heights of $\sim100\,\mathrm{kpc}$ above the disc. \citet{EverettSchillerZweibel2010} also find increasing wind speeds as a function of height with wind speeds of $v>300\,\mathrm{km\,s}^{-1}$ at heights within $1\,\mathrm{kpc}$ from the disc. However, their model assumes all the CRs to be injected at $z=0$, whereas our simulations use vertical SN distributions -- and thus CR sources -- up to a height of $1\,\mathrm{kpc}$. The three-dimensional models by \citet{HanaszEtAl2013} and \citet{PakmorEtAl2016} also find increasing velocity profiles as a function of height with outflowing velocities below $100\,\mathrm{km\,s}^{-1}$ close to the midplane.

\section{Conclusions}

We perform magnetohydrodynamical simulations of the SN- and CR-driven ISM in stratified boxes using \textsc{FLASH}. We include a chemical network that follows the abundances of ionised, atomic and molecular hydrogen, CO and C$^+$. We assume an interstellar radiation field that is locally attenuated in optically thick regions. The shielding of the radiation is computed using the TreeCol algorithm. SNe are injected at a constant rate and are spatially clustered. When the first SN of a cluster explodes we place the cluster in the largest density peak where it remains for the rest of its lifetime. CRs are included as a separate relativistic fluid in the advection-diffusion approximation. The CR pressure is added to the total pressure and therefore is dynamically coupled with the gas. The diffusion is treated in an anisotropic way with large diffusion coefficients along the magnetic field lines ($1-3\times10^{28}\,\mathrm{cm}^2\,\mathrm{s}^{-1}$) and two orders of magnitude smaller values perpendicular to the local magnetic field. In two simulations we additionally couple the locally varying CR energy density to the checmical network via the CR ionisation rate. In the other simulations the ionisation rate is constant in space and over time. Our results can be summarised as follows:

\begin{itemize}
\item CRs help driving outflows. The outflow rates strongly vary over time with temporally indistinguishable rates for runs including and excluding CRs. However, over time the CRs support the SNe efficiently enough to drive twice as much mass out of the disc into the halo. The overall outflow rates are of order the star formation rate, i.e. at mass-loading factors of order unity. The CR supported outflows are denser by almost an order of magnitude ($\rho\sim10^{-26}\,\mathrm{g~cm}^{-3}$) which outweighs the smaller outflow velocities (factor of $2-3$ with $v_z\sim30-40\,\mathrm{km~s}^{-1}$) to give a net higher mass loading factor. Computing an effective mass loading factor based on all the mass above a height of $2.5\,\mathrm{kpc}$ over the last $100\,\mathrm{Myr}$ of evolution yields $\eta_\mathrm{th}\sim0.1$ for the thermal run and $\eta_\mathrm{CR}\sim0.7-1.4$ for the simulations including CRs.

\item SNe strongly shape their local environment. Up to altitudes of $\sim1\,\mathrm{kpc}$ the gas is highly structured with high clumping factors ($C_\rho\sim100$) and large volume filling fractions of the hot ($T>3\times10^5\,\mathrm{K}$) gas, $\mathrm{VFF_\mathrm{hot}}\gtrsim0.6$. Above the region with direct SN impact ($z\gtrsim1\,\mathrm{kpc}$) the CR supported outflows form a smooth gas distribution with smaller and temporally more steady clumping factors ($C_\rho=2-5$ vs. $C_\rho=1-50$). In addition, the gas is colder ($T\sim10^4\,\mathrm{K}$ vs. $T\sim2\times10^6\,\mathrm{K}$ in thermal outflows), which reduces the volume filling fraction of hot gas ($T>3\times10^5\,\mathrm{K}$) from $>90\%$ (thermal run) to $\lesssim20\%$ (CR runs).

\item If the SNe are resolved and partially explode in low-density regions they are able to temporarily push gas to heights of $1-2.5\,\mathrm{kpc}$ at mass-loading factors of order unity. This emphasises the importance of accurately resolving the SNe as well as placing the SNe at the appropriate positions in the disc, in particular because the CRs mainly accelerate the diffuse gas rather than the gas bound to molecular clouds. CRs are able to efficiently support the SN-driven outflows. Whereas the simulation with purely thermal SN energy injection develops fountain flows that partially start falling back towards the disc after $\sim100\,\mathrm{Myr}$, the corresponding CR-supported outflows reach a steady state for intermediate CR diffusion coefficients ($\mathsf{K}_\parallel=3\times10^{28}\,\mathrm{cm}^2\,\mathrm{s}^{-1}$) and even increase the mass-loading in the case of smaller diffusion coefficients ($\mathsf{K}_\parallel=10^{28}\,\mathrm{cm}^2\,\mathrm{s}^{-1}$).

\item The vertical acceleration away from the disc due to thermal and CR pressure gradients can compensate the gravitational attraction towards the midplane above a height of approximately $0.5-1\,\mathrm{kpc}$. For $\mathsf{K}_\parallel=3\times10^{28}\,\mathrm{cm}^2\,\mathrm{s}^{-1}$ the gas effectively moves in a force-free manner. For $\mathsf{K}_\parallel=10^{28}\,\mathrm{cm}^2\,\mathrm{s}^{-1}$ the net acceleration points outwards and can further accelerate the outflow.

\item The CR energy density is relatively smoothly distributed in the box with a mean of $e_{_\mathrm{CR}}\sim3-5\times10^{-12}\,\mathrm{erg\,cm}^{-3}$ and a width of less than an order or magnitude. Local variations quickly reduce because of fast diffusion. Locally on scales of a few 100 pc there is no correlation between CR energy and the gas. However, on scales of the entire box there is a small positive correlation, which manifests in $\gamma$-ray clumping factors of $2-6$.

\item The disc ($|z|<0.1\,\mathrm{kpc}$) shows strong thermal pressure variations over a range of more than four orders of magnitude. As the CR pressure only varies by less than one order of magnitude, the range of $X_\mathrm{CR}=P_\mathrm{CR}/P_\mathrm{therm}$ locally ranges from $0.01-100$. Averaged over the volume the disc is dominated by CR pressure with $X_\mathrm{CR}$ in the range from $1$ to $10$. In the lower halo ($0.1\,\mathrm{kpc}<|z|\le1\,\mathrm{kpc}$) $X_\mathrm{CR}$ is reduced because SNe can efficiently heat the gas to temperatures above $10^6\,\mathrm{K}$ resulting in very high thermal pressures. Above an altitude of $1\,\mathrm{kpc}$ there is no direct heating by SNe and the CR pressure dominates over the thermal counterpart with $X_\mathrm{CR}$ exceeding $100$.

\item Approximately $5-25\%$ of the injected CR energy cools via hadronic losses. As the hadronic losses scale with $\rho e_{_\mathrm{CR}}$ and the CR energy density is relatively smooth, the majority of the losses occurs in the disc region, where the density is highest. For smaller diffusion coefficients the CRs escape from the dense gas on longer time scales, which results in more efficient cooling.

\item An active CR ionisation rate does not have a significant impact on the dynamics on scales above a few hundred parsecs. Local changes in the density close to the SNe affect the subsequent SNe and temporarily change the dynamics. However, there is no systematic difference between a constant and a locally varying CR ionisation rate in the halo gas distribution, the outflows and pressure ratios.

\end{itemize}

\section*{Acknowledgements}
We thank Christoph Pfrommer, Andrew Strong, the members of the SILCC collaboration, Mordecai-Mark Mac Low, R\"{u}diger Pakmor and Christine Simpson for stimulating discussions. We also thank the anonymous referee for a critical reading and valuable suggestions that improved the clarity of the paper. P.G., S.W., and T.N. acknowledge support from the DFG Priority Program 1573 {\em Physics of the Interstellar Medium}.
P.G. acknowledges funding from the European Research Council under ERC-CoG grant CRAGSMAN-646955.
T.N. acknowledges support from the DFG cluster of excellence \emph{Origin and Structure of the Universe}.
M.H. acknowledges support of the (Polish) National Science Centre through the grant No. 2015/19/ST9/02959.
S.W. acknowledges the support of the Bonn-Cologne Graduate School, which is funded through the Excellence Initiative. S.W. further acknowledges the support from the European Research Council through the ERC Starting Grant RADFEEDBACK (project number 679852) under FP8.
The authors thank the Max Planck Computing and Data Facility (MPCDF) for computing time and data storage.
The software used in this work was developed in part by the DOE NNSA ASC- and DOE Office of Science ASCR-supported Flash Center for Computational Science at the University of Chicago. Parts of the analysis are carried out using the \textsc{YT} analysis package (\citealt{TurkEtAl2011}, \url{yt-project.org}).

%%%%%%%%%%%%%%%%%%%%%%%%%%%%%%%%%%%%%%%%%%%%%%%%%%



\bibliographystyle{mnras}
\bibliography{astro} % if your bibtex file is called example.bib

%%%%%%%%%%%%%%%%% APPENDICES %%%%%%%%%%%%%%%%%%%%%
\appendix

\section{Self-gravity or not?}
\label{sec:app-self-gravity}

\begin{figure*}
\begin{minipage}{\textwidth}
\includegraphics[height=0.45\textheight]{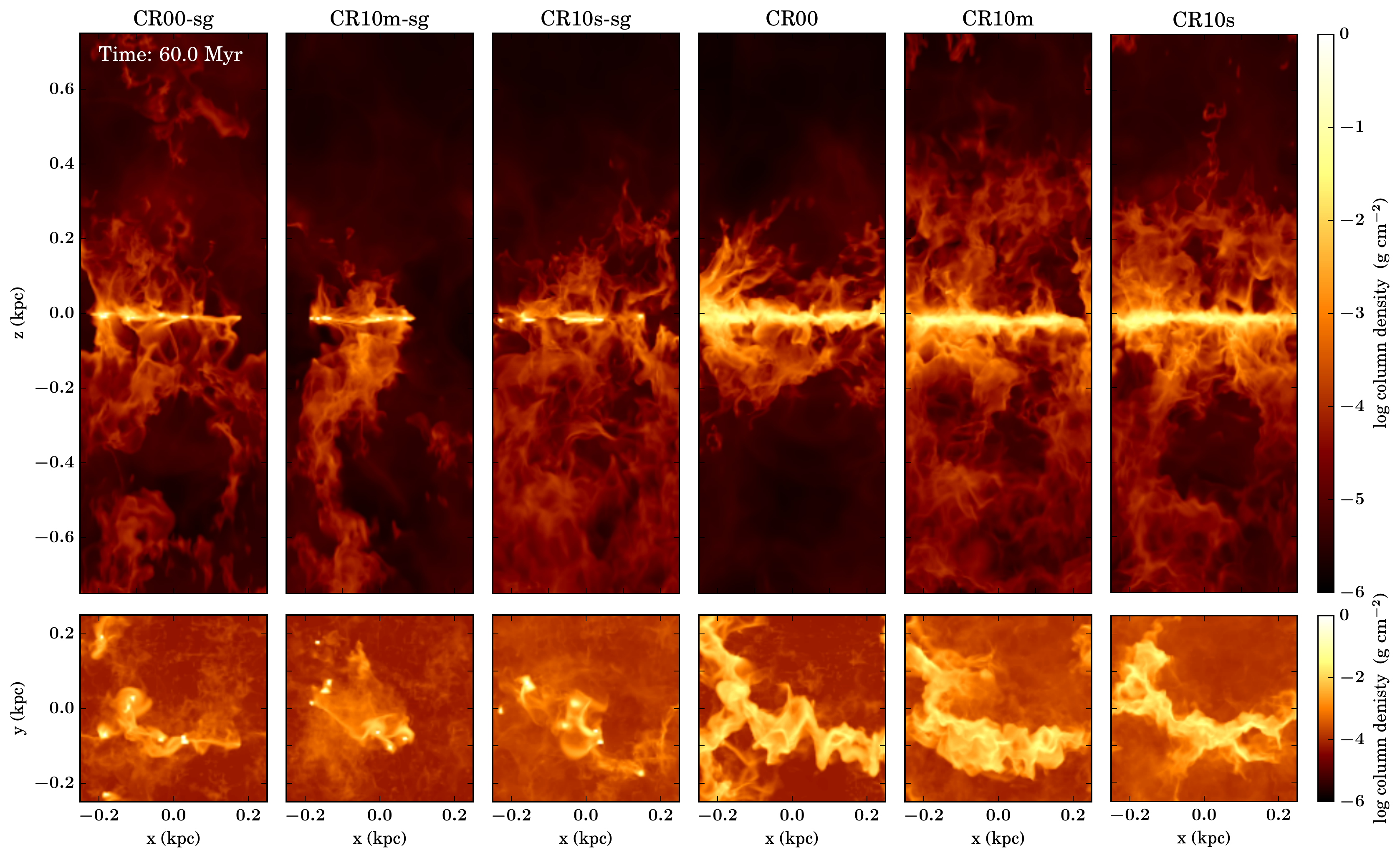}
\end{minipage}
\caption{Column density plots for all simulations at $60\,\mathrm{Myr}$. The top panels show the egde-on view, the bottom panels the face-on view. The left three pictures are for the simualtions (\texttt{noCR-sg}, \texttt{CR-medK-sg}, \texttt{CR-medK-loc}$\zeta_\mathrm{CR}$\texttt{-sg}) with self-gravity, the three on the right are for the corresponding non-selfgravity runs. After $t=60\,\mathrm{Myr}$ self-gravity has accumulated the vast majority of the gas in a few dense clouds. The differences between CR and no-CR simulations is small in this case. Without self-gravity the disc is thicker and the differences between CR and no-CR is apparent.}
\label{fig:coldens-comparison}
\end{figure*}

On the one hand self-gravity is an important process for the structures in the ISM, in particular the dense regions in the disc. On the other hand a numerical experiment is not possible if self-gravity dominates the dynamical evolution and accumulates all gas in clouds that cannot be dispersed by the implemented feedback processes. As discussed in detail in \citet{GirichidisSILCC2} and Girichidis et al. (2018, SILCC-V, submitted) the SN feedback in our setups does not prevent the gas structures to continuously grow in mass in the presence of self-gravity. Fig.~\ref{fig:coldens-comparison} shows an edge-on (elongated) and face-on (square) projection of the density at $t=60\,\mathrm{Myr}$. The left three simulations are the ones including self-gravity, the three on the right are without self-gravity. All simulations including self-gravity show dense clouds that contain most of the mass. The differences between with and without CRs is very small because the gravitational attraction is significantly stronger than the thermal and CR pressure gradients. As most of the gas is locked up in the dense clumps there is no diffuse gas available for outflows. For short simulation times of the order of a few tens of Myr the problem might not be severe. However, we would like to follow the evolution for many gravitational collapse times and need to assure that the gas is not unrealistically collapsed into a few compact objects.

\section{Resolution effects}
\label{sec:app-resolution}

% Hill plots
\begin{figure*}
\begin{minipage}{\textwidth}
\includegraphics[height=0.45\textheight]{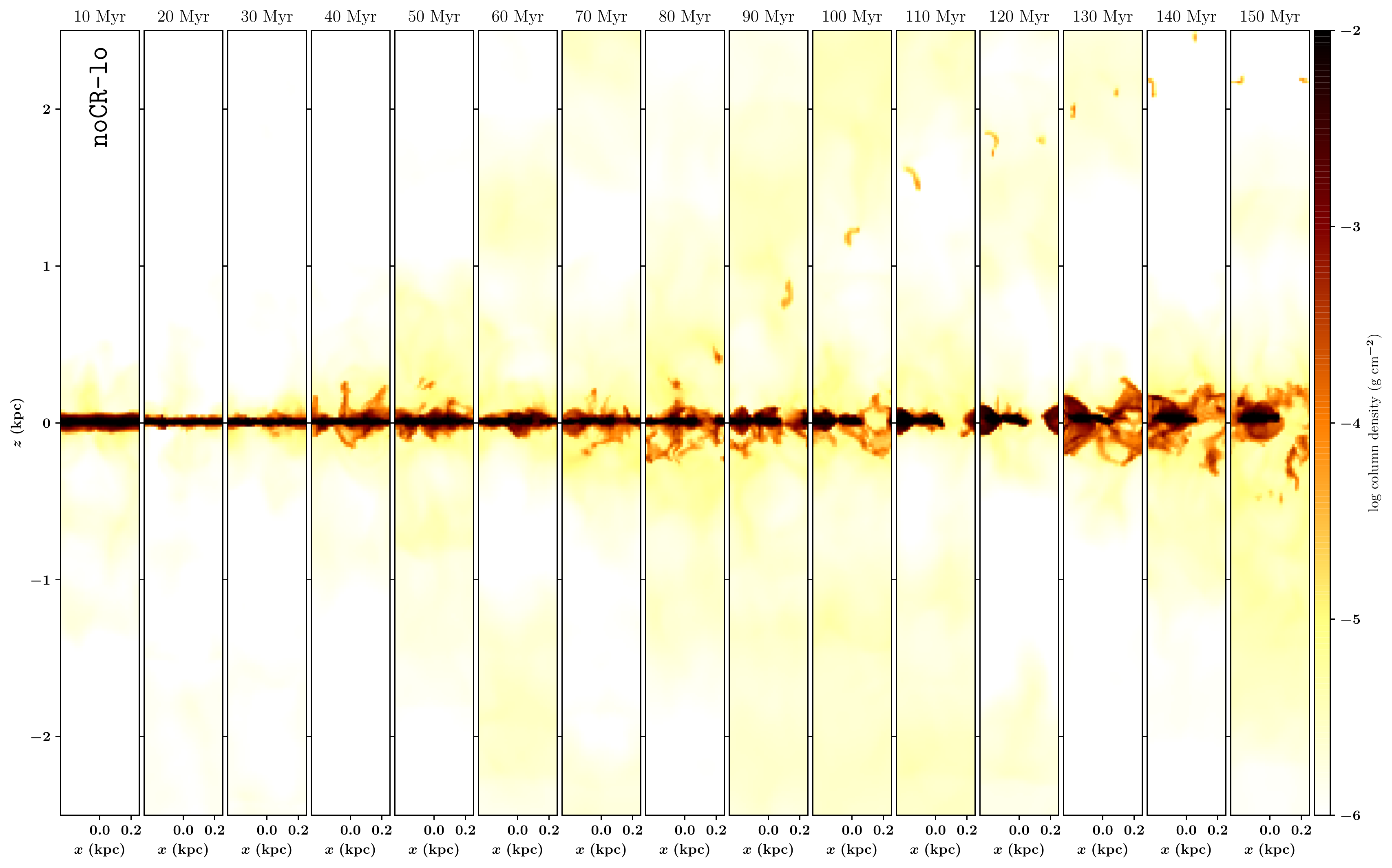}\\
\includegraphics[height=0.45\textheight]{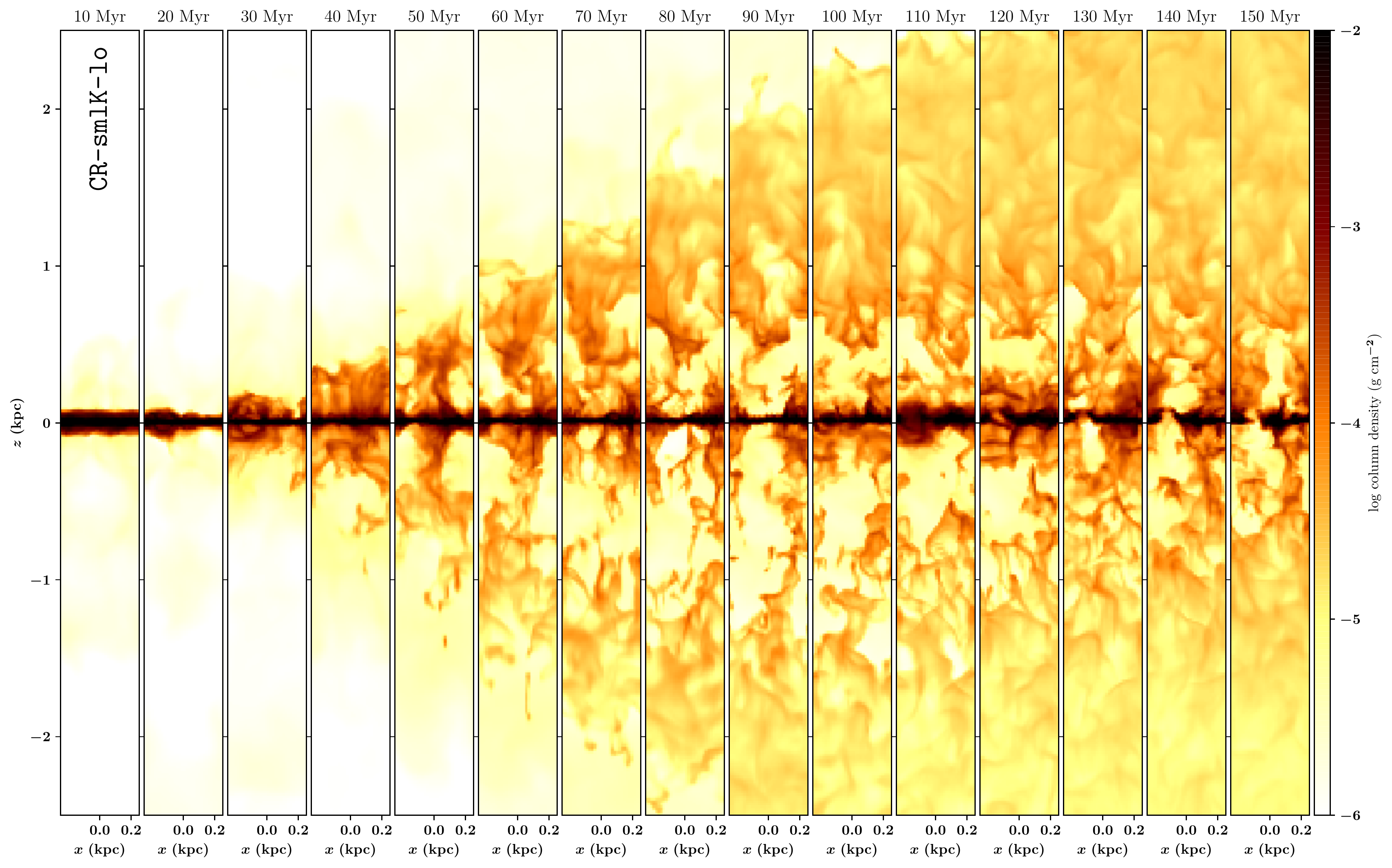}
\end{minipage}
\caption{Time evolution for simualtions \texttt{noCR-lo} (top) and \texttt{CR-smlK-lo} (bottom), shown is the column density edge-on in steps of $10\,\mathrm{Myr}$. The purely thermal run creates patchy outflows with large voids in the halo. Towards the end of the simulation the gas begins to fall back towards the disc. The CR run in the bottom launches volume filling outflows. Up to a height of $1\,\mathrm{kpc}$ the direct SN impact creates a structured medium. Above that height the gas redistributes with low column-density contrasts.}
\label{fig:Hill-plots-00-10s-ng-lowres}
\end{figure*}

\begin{figure*}
\begin{minipage}{\textwidth}
\includegraphics[height=0.45\textheight]{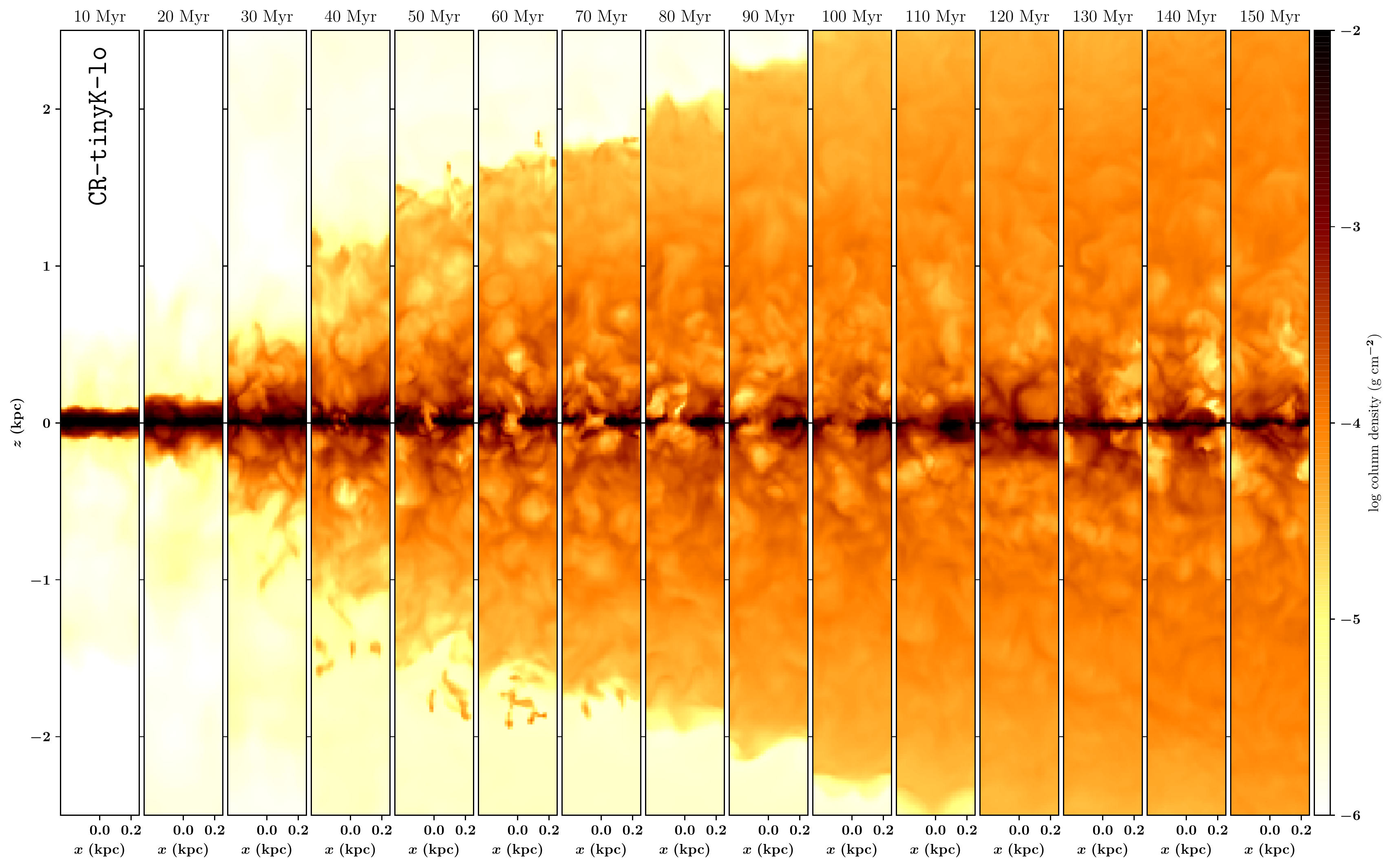}\\
\includegraphics[height=0.45\textheight]{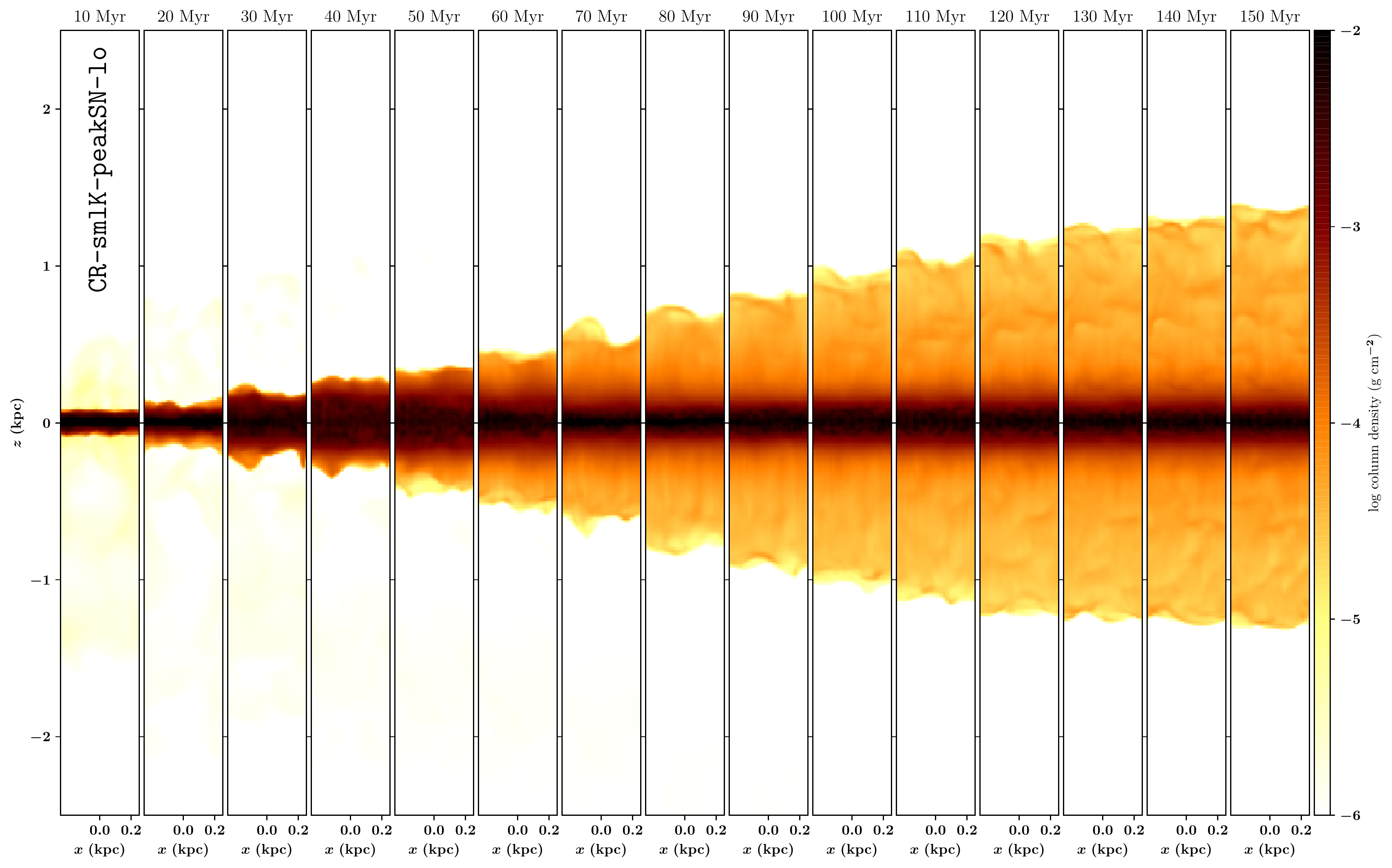}
\end{minipage}
\caption{Time evolution for simualtions \texttt{CR-tinyK-lo} (top) and \texttt{CR-smlK-peakSN-lo} (bottom), shown is the column density edge-on in steps of $10\,\mathrm{Myr}$. Both simulations show similar features in terms of the structure of the gas and the lauching of outflows. However, the run with locally varying CR ionisation rate needs more time to lift the gas to heigths of $\sim1\,\mathrm{kpc}$.}
\label{fig:Hill-plots-10u-10sp-ng-lowres}
\end{figure*}

\begin{figure*}
\begin{minipage}{\textwidth}
\includegraphics[width=\textwidth]{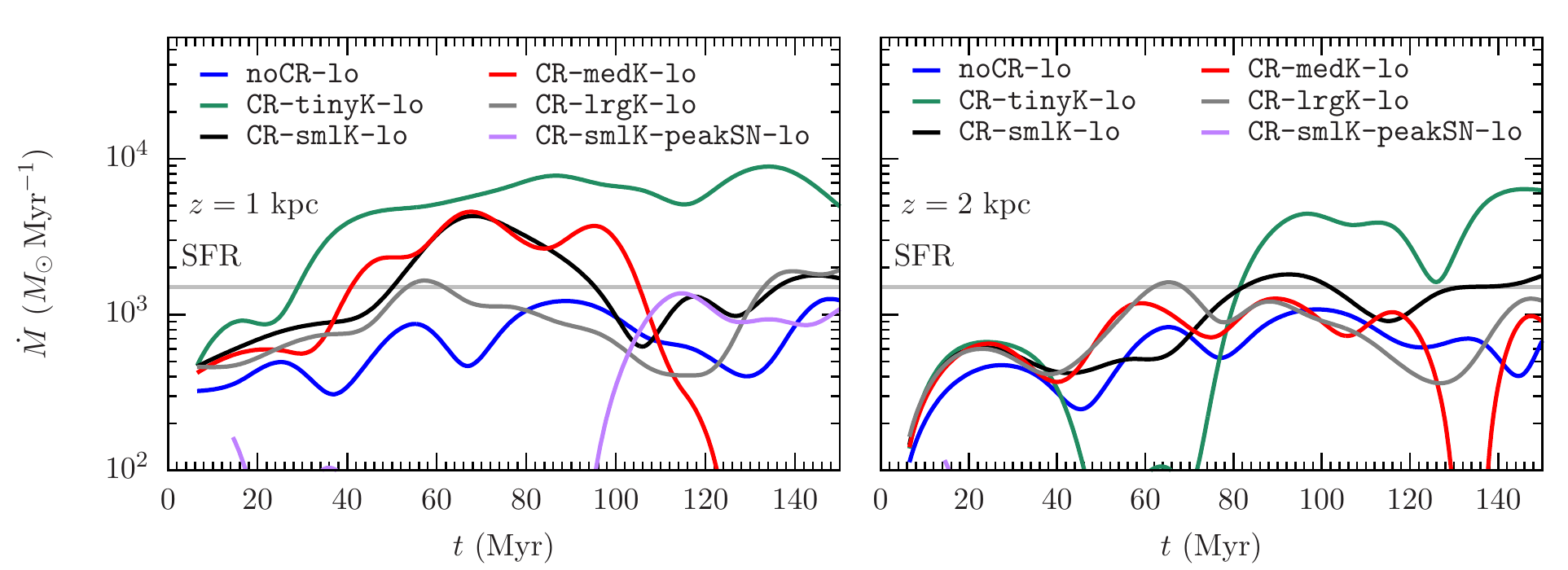}
\end{minipage}
\caption{Outflow rates for low-resolution simulations. The effect of the SNe is reduces because their Sedov-Taylor radius is not resolved. The CRs can drive outflows at a rate comparable to the star formation rate ($\eta\sim1$).}
\label{fig:outflow-time-evol-lowres}
\end{figure*}

\begin{figure}
\includegraphics[width=8cm]{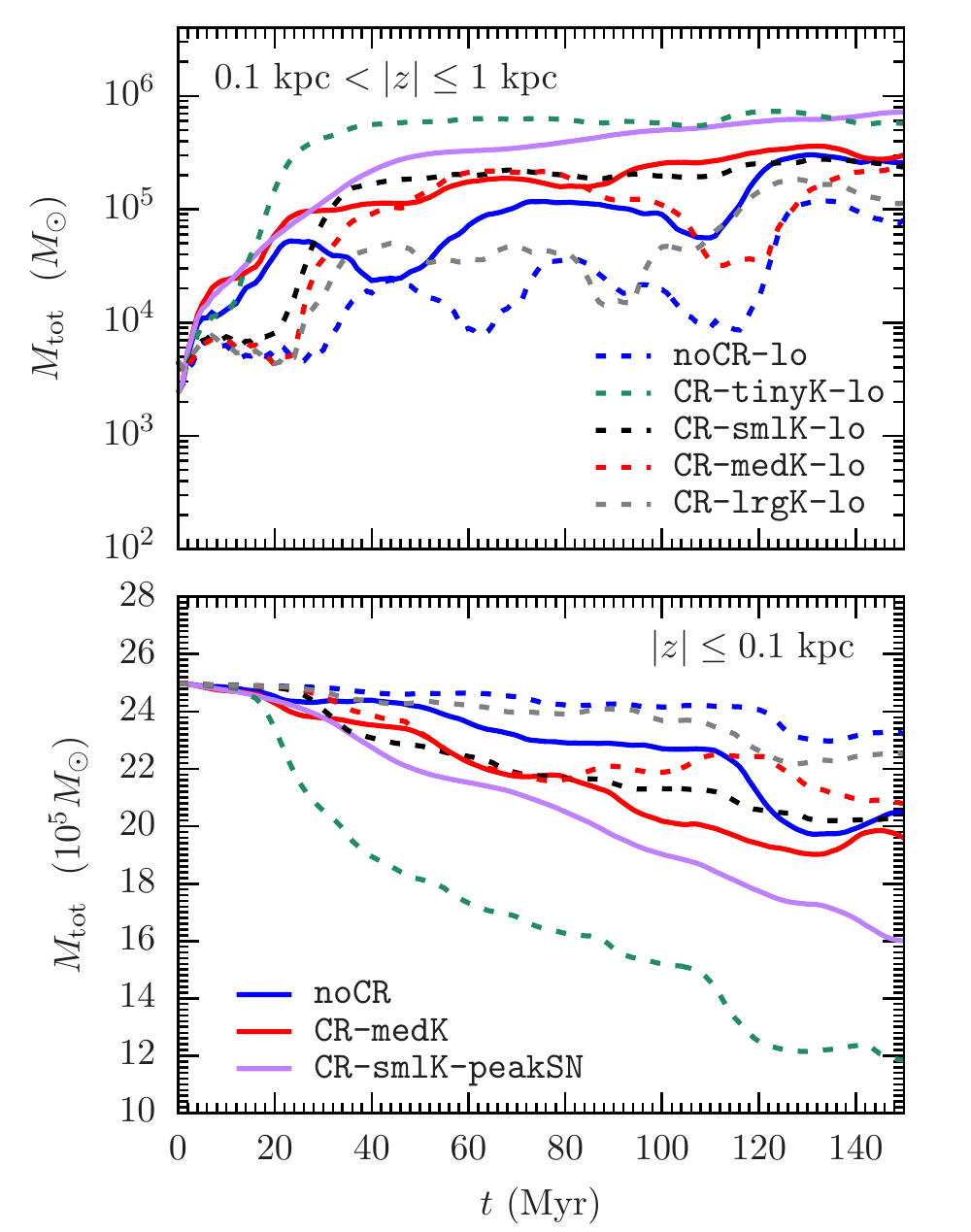}
\caption{Halo masses at different vertical heights as a function of time for different resolutions and different diffusion coefficients. The lower the diffusion coefficient the more gas is transported out of the disc into the halo.}
\label{fig:halo-mass-diff-heights-lowres}
\end{figure}

For the SNe we would like to resolve the Sedov-Taylor radius with at least 4 cells in order to properly account for their thermal impact. The gradient of the CR pressure varies on larger scales, which suggests that the impact of CRs can be captured with lower resolution. We probe the dependence of the CR support for driving winds by lowering the resolution by a factor of 4. This results in a smallest cell size of $16\,\mathrm{pc}$. The SN rate as well as the fractions of individual, clustered and type~Ia SNe is the same as in the high-resolution runs. The positioning of the clusters is determined from the densest point in the simulation box at the time a new cluster is activated. This means that the positions of the clusters vary. For the lower resolution the Sedov-Taylor radius is not resolved with 4 cells, which then underestimates the thermal impact of the SNe.

The time evolution of the column density is shown in Fig.~\ref{fig:Hill-plots-00-10s-ng-lowres} for simulations \texttt{noCR-lo} (top) and \texttt{CR-smlK-lo} (bottom) and Fig.~\ref{fig:Hill-plots-10u-10sp-ng-lowres} for \texttt{CR-tinyK-lo} (top) and \texttt{CR-smlK-peakSN-lo} (bottom). The strong thermal losses in \texttt{noCR-lo} results in perturbations in the disc but does not lead to a noticeable outflow. This is consistent with the result in \citet{GirichidisEtAl2016CR}. Patches of outflowing material are visible, but the column densities of the gas in the halo are more than an order of magnitude lower than in the high resolution run. The time evolution of run \texttt{CR-lrgK-lo} with a large CR diffusion coefficient ($\mathsf{K}_\parallel=10^{29}\,\mathrm{cm}^2\,\mathrm{s}^{-1}$) looks very similar to the purely thermal low-resolution run (not shown). This is not surprising because the fast diffusion removes the CR pressure gradients before they can accelerate the gas and launch an outflow. For lower diffusion coefficients ($\mathsf{K}_\parallel=10^{27}\,\mathrm{cm}^2\,\mathrm{s}^{-1}$, \texttt{CR-smlK-lo}, Fig.~\ref{fig:Hill-plots-00-10s-ng-lowres} bottom) we note similar features as in the high-resolution run. Gas is lifted out of the disc and fills the halo without a sign of turnover. The SNe that explode in the lower halo ($0.1\,\mathrm{kpc}<|z|\le1\,\mathrm{kpc}$) reshape the gas with local column density contrasts of $1-2$ orders of magnitude. Above the direct impact of SNe the gas becomes smooth. As the SNe are poorly resolved the CRs are effectively the main driving mechanism to launch the outflow. Lowering the diffusion coefficient to $\mathsf{K}_\parallel=10^{28}\,\mathrm{cm}^2\,\mathrm{s}^{-1}$ (\texttt{CR-tinyK-lo}, Fig.~\ref{fig:Hill-plots-10u-10sp-ng-lowres} top) aggravates the impact of CRs. The slow diffusion retains larger CR pressure gradients. The time scales for hadronic losses are larger than the dynamical time scales for launching an outflow, so the halo is filled with dense gas that is lifted from the disc. The density in the lower halo is high enough for the thermal SN energy to be quickly radiated away. As a result, the column density contrast is very low and the transition between the lower and upper halo is almost invisible. Placing all SNe in density peaks (\texttt{CR-smlK-peakSN-lo}, Fig.~\ref{fig:Hill-plots-10u-10sp-ng-lowres} bottom) leads to a strong thermal overcooling and no relevant density perturbations. The disc is thicker than in the other runs, where a significant fraction of the SNe explode in clusters or outside the disc region, and the CRs are able to lift dense and smooth gas into the halo. However, the gas needs almost twice as long to reach a certain height compared to the high-resolution run \texttt{CR-smlK-peakSN}.

Fig.~\ref{fig:outflow-time-evol-lowres} shows the outflow rates for the low-resolution runs. Plotted is the net outflow rate,
\begin{equation}
\dot{M} = \sum_i\rho_i\,\mathrm{sgn}(z)\, v_{z,i}\,A_i,
\end{equation}
not split into \emph{inflow} and \emph{outflow} as in the high resolution case. Here, $\rho_i$, $v_{z,i}$, and $A_i=dx_i\times dy_i$ are the individual cell densities, $z$ components of the velocities and surface areas. The set of cells, $i$, includes all cells that intersect with the measurement positions at $\pm1\,\mathrm{kpc}$ (left) and $\pm2\,\mathrm{kpc}$ (right), respectively. Overall the outflow properties are similar despite the smaller thermal impact of the SNe. In the case of purely thermal energy injection the outflow rates are always below the star formation rate, i.e. at $\eta<1$. This is different from the run with $4\,\mathrm{pc}$ resolution, where the thermal energy of the SNe alone was able to reach outflow rates above unit, even though only for a limited time of $\sim100\,\mathrm{Myr}$ before the gas starts falling back in a fountain flow. The runs including CRs are lacking the thermal support from the under-resolved SNe, which results in smaller outflow rates. Nonetheless, the they reach rates with $\eta\sim1-8$ for a relevant fraction of the simulation time. The temporal fluctuations are large but there is still a noticeable trend in the mass loading factor with the CR diffusion coefficient. Run \texttt{CR-lrgK-lo} drives a continuous outflow at $\eta<1$, runs \texttt{CR-medK-lo} and \texttt{CR-smlK-lo} exceed $\eta=1$ and the strongest outflows with $\eta\sim5$ are launched in \texttt{CR-tinyK-lo}. The peak driving run needs the majority of the simulation time to lift gas to a height of $1\,\mathrm{kpc}$ but then continues to dive gas through this measurement altitude at a rate of $\eta\sim1$. The under-resolved thermal effect of the SNe emphasises that CRs alone are able to drive outflows with mass loading factors of order unity.

As the outflows show strong fluctuations over time we also plot the mass in the disc ($|z|<0.1\,\mathrm{kpc}$) and the lower halo ($0.1\,\mathrm{kpc}\le|z|<1\,\mathrm{kpc}$) as a function of time in Fig.~\ref{fig:halo-mass-diff-heights-lowres}. The dashed lines are for the low-resolution simulations, the solid lines indicate the high-resolution counterpart. The strong trend of the amount of outflowing gas with the diffusion coefficient is evident in both panels. For \texttt{CR-tinyK-lo} the disc experiences the strongest mass loss by far. Simulations \texttt{CR-medK-lo} and \texttt{CR-smlK-lo} are comparable to their high-resolution counterpart with slightly more mass in the disc towards the end of the simulations, which emphasises the importance of resolving the SNe. Simulation \texttt{CR-lrgK-lo} is comparable to the purely thermal run \texttt{noCR-lo} with very little outflows. The masses in the lower halo inversely reflect the same trends. Changing the diffusion coefficient from $10^{27}$ to $10^{29}\,\mathrm{cm}^2\mathrm{s}^{-1}$ results in a decrease of the lower halo mass by approximately one order of magnitude.

% Don't change these lines
\bsp	% typesetting comment
\label{lastpage}
\end{NoHyper}
\end{document}